\documentclass{aa}

\usepackage{graphicx}
\usepackage{xcolor}
\usepackage{txfonts}
\usepackage[breaklinks=true]{hyperref}
\usepackage{placeins}

\begin{document} 

   \title{The eROSITA view on the halo mass–temperature relation:  \\ From low-mass groups to massive clusters}

   \author{V. Toptun\inst{1}\thanks{victoria.toptun@eso.org}
   \and
        P. Popesso\inst{1,}\inst{2}
          \and
          I. Marini\inst{1}
          \and
          K. Dolag\inst{3,}\inst{4,}\inst{2}
          \and
          G. Lamer\inst{5}
          \and
          X. Yang\inst{6}
          \and
          Q. Li\inst{6}
          \and
          B. Csizi\inst{7}
          \and
          L. Lovisari\inst{8}
          \and
          S. Ettori\inst{9,}\inst{10}
          \and
        V. Biffi\inst{11,}\inst{12}
          \and
            S. Vladutescu-Zopp\inst{3}
          \and
          A. Dev\inst{13}
          \and
          D. Mazengo\inst{1,}\inst{17}
          \and
          A. Merloni\inst{14}
          \and
          J. Comparat\inst{14}
          \and
          G. Ponti\inst{15,}\inst{14,}\inst{16}
          \and
          E. Bulbul\inst{14}
          }

   \institute{European Southern Observatory, Karl Schwarzschildstrasse 2, 85748, Garching bei M\"unchen, Germany
         \and
            Excellence Cluster ORIGINS, Boltzmannstr. 2, D-85748 Garching bei M\"unchen, Germany
        \and
             Universitäts-Sternwarte, Fakultät für Physik, Ludwig-Maximilians-Universität München, Scheinerstr.1, 81679 München, Germany
        \and 
            Max-Planck-Institut für Astrophysik, Karl-Schwarzschildstr. 1, 85741 Garching bei M\"unchen, Germany
        \and
            Leibniz-Institut für Astrophysik Potsdam (AIP), An der Sternwarte 16, 14482 Potsdam, Germany  
        \and
            Shanghai Astronomical Observatory (SHAO) at the Chinese Academy of Sciences, 80 Nandan Road, Xuhui District, Shanghai 200030, China
        \and
        Universität Innsbruck, Institut für Astro- und Teilchenphysik, Technikerstr. 25/8, 6020 Innsbruck, Austria
                \and
            INAF, Istituto di Astrofisica Spaziale e Fisica Cosmica di Milano, via A. Corti 12, 20133 Milano, Italy
            \and
            INFN, Sezione di Bologna, viale Berti Pichat 6/2, 40127 Bologna, Italy 
        \and
            INAF, Osservatorio di Astrofisica e Scienza dello Spazio, via Piero Gobetti 93/3, 40129 Bologna, Italy
        \and
             INAF, Osservatorio Astronomico di Trieste, Via Tiepolo 11, 34143 Trieste, Italy
                     \and 
            IFPU, Institute for Fundamental Physics of the Universe, Via Beirut 2, I-34014 Trieste, Italy
            \and
            International Centre for Radio Astronomy Research, University of Western Australia, M468, 35 Stirling Highway, Perth, WA 6009, Australia
                \and
            Max-Planck-Institut für Extraterrestrische Physik (MPE), Giessenbachstr. 1, D-85748 Garching bei München, Germany
                    \and
            INAF– Osservatorio Astronomico di Brera, Via E. Bianchi 46, 23807 Merate (LC), Italy
            \and
            Como Lake Center for Astrophysics (CLAP), DiSAT, Universit\`a degli Studi dell’Insubria, via Valleggio 11, I-22100 Como, Italy
            \and
            Physics Department, College of Natural and Mathematical Sciences, P.O.Box 338, The University of Dodoma, Tanzania.
             }

   \date{Received ; accepted }
 
  \abstract
    {Galaxy groups and clusters are among the best probes of structure formation and growth in a cosmological context. 
     Most of their baryonic component is dominated by hot plasma, known as the intracluster medium (ICM) in clusters or the intragroup medium (IGrM) in groups. Their thermodynamical properties serve as indicators of the halo's dynamical state and can be used to determine halo mass in the self-similar scenario. However, baryonic processes, such as AGN feedback and gas cooling, may affect the global properties of the ICM, especially in the group regime. These effects might lead to deviations from self-similar predictions in the scaling relations of galaxy groups, while they remain in place for massive galaxy clusters. Additionally, the low-mass end of the scaling relations, ranging from $10^{13}$ to $10^{14} M_\odot$, remains unclear and poorly populated, as current X-ray surveys detect only the brightest groups. Here, we present the mass-temperature relation across the entire mass range, from massive clusters to low-mass groups ($10^{13}M_\odot$), as observed by eROSITA. Using spectral stacking from the first eROSITA All-Sky Survey data for optically selected galaxy groups, we find that, in the lower mass range, galaxy groups follow the power-law relation known for galaxy clusters. We provide the best-fit mass–temperature relation, validated over two decades in halo mass, as follows:
  $\log_{10} (M_{500}/M_{\odot}) = (1.65\pm0.11)\cdot \log_{10} ({T_{X}}/{1\;\mathrm{keV}}) + (13.38\pm0.05)$.
  We further validate these results by conducting the same stacking procedure on mock eRASS:4 data using the {\sc Magneticum} hydrodynamical simulation. This indicates that AGN feedback is more likely to affect the distribution of baryons in the intragroup medium rather than the overall halo gas temperature. No significant changes in the slope of the mass-temperature relation suggest that temperature can serve as a reliable mass proxy across the entire mass range. This supports the use of temperature-derived masses, particularly in cosmological studies, significantly broadening the mass range and enabling applications such as improving cluster mass function studies and cosmological parameter estimates.}

   \keywords{Galaxies: groups: general - X-rays: general - X-rays: galaxies: clusters - galaxies: active - methods: data analysis
               }

   \maketitle
   
\section{Introduction}

In a pure gravitational picture, the temperature and density distribution of the halo's hot gas, and thus its X-ray emission, are dictated by the halo's potential well, which follows the dark matter (DM) distribution profile. Under these conditions, the slope of the power-law relation between the X-ray luminosity ($L_X$) of the intragroup and intracluster medium (IGrM and ICM, respectively) and the halo mass ($M_{halo}$) is predicted to be 4/3 for the bolometric X-ray and 1 for the soft-band X-ray in the regime of self-similarity, as inherited by massive halos during the evolution of the large-scale matter distribution. Correspondingly, the self-similar model predicts $M \propto T^{1.5}$ scaling for the temperature-mass relation {(e.g., \citealt{self_sim, 2013SSRv..177..247G, 2022hxga.book...65L})}. However, recent X-ray observations indicate substantial deviations from this expected self-similarity, revealing relations of $L_X \propto T^{2.7-3.0}$ and $M \propto T^{1.6-1.8}$ {\citep{2016MNRAS.463.3582M,2019ApJ...871...50B,2020ApJ...892..102L,2022MNRAS.511.4991A}}. Moreover, these discrepancies become more pronounced as the mass scale of cosmic structures decreases.

The divergence between self-similar predictions and observations is likely attributable to nongravitational processes occurring at the cores of clusters, groups, and galaxies, including AGN feedback, shock heating, supernova heating, and gas cooling (e.g., \citealt{2007MNRAS.380..877S, 2008ApJ...687L..53P, fabjan_simulating_2010, 2010MNRAS.406..822M, 2014MNRAS.441.1270L}). The energy supplied by supernovae and active galactic nuclei (AGNs) to the hot IGrM can easily exceed the gravitational binding energy of less massive groups, whereas in clusters, feedback effects are typically confined to the central core {\citep{2014MNRAS.441.1270L,2021Univ....7..142E}}. Consequently, nongravitational mechanisms are expected to significantly influence baryons within group halos, making them ideal targets for constraining the mechanisms that govern the cooling–heating balance \citep{pillepich}. By extending studies of X-ray scaling relations from clusters to low-mass faint groups, we can simultaneously probe the effects of nongravitational heating processes on large-scale structure and galaxy evolution. This paper investigates how these phenomena impact the $M-T$ relation across a broad range of temperature and mass scales.

Previous measurements of the $M-T$ relation have primarily concentrated on galaxy clusters and X-ray-bright groups {\citep{2009ApJ...692.1033V, 2014Ap&SS.349..415B, kettula_mt, 2019ApJ...871...50B, 2022A&A...661A...7B}}. It is established that the $M-T$ scaling relation for nearby clusters is less influenced by nongravitational processes than other scaling relations, exhibiting smaller deviations from self-similarity. The $M-T$ relation for the most massive galaxy clusters, predominantly those with $T > 2$ keV, follows a power law with a slope of 1.6$-$1.8 \citep{2000ApJ...532..694N, 2000ApJ...542..106T, 2001A&A...368..749F, 2009ApJ...692.1033V, 2013arXiv1302.0873B, 2014Ap&SS.349..415B}. At lower masses, the availability of homogeneously selected massive X-ray groups remains limited across various depths and resolutions \citep{1996MNRAS.283..690P, 2000ARA&A..38..289M, 2004MNRAS.350.1511O, sun_mt, eckmiller_mt, kettula_mt, lovisari_relation, umetsu_mt}. The challenge of probing the $M-T$ relation in the majority of the group population has so far been hampered by the insufficient sensitivity of previous X-ray surveys. 

The eROSITA All Sky Survey (eRASS), which began in December 2019, is expected to yield the largest X-ray catalog of galaxy groups in the local Universe upon its completion {\citep{2021A&A...647A...1P, 2021A&A...656A.132S, 2024A&A...685A.106B}}. Nevertheless, the data corresponding from the first-year scan (eRASS1, \citealt{2024A&A...682A..34M}) remain too shallow to provide insights into the bulk of the local galaxy group population through individual observations. Mock eROSITA observations derived from the {\sc Magneticum} magnetohydrodynamical simulation indicate that, at the depth of eRASS:4 (second-year scanning, cumulative result of four consecutive stacked surveys), eROSITA will detect only a marginal fraction of the group population at $z<0.2$, particularly excluding groups with high core entropy and lower gas fractions \citep{ilaria_lightcone, paola_stacking_magneticum, paola_gasfraction}. This is consistent with analyses of the eROSITA selection function \citep{2024A&A...687A.238C, 2024A&A...686A.196S} and suggests that the X-ray selection process itself may introduce biases in the mean properties of groups \citep{paola_24}. Specifically, if AGN feedback is capable of removing portions of the IGrM, this could significantly skew the X-ray selection. Thus, studying the hot gas properties in galaxy groups necessitates selecting systems through a method that is unbiased by such feedback and independent of the hot gas characteristics.

The most straightforward approach to achieve this is through optical selection, which is based on the clustering properties of the spectroscopic galaxy population according to their redshifts and luminosities. \cite{ilaria_opticallightcone} tested and compared the most widely used spectroscopic optical selection algorithms \citep{robotham, yang_galaxy_2007, tempel} on galaxy lightcones derived from the {\sc Magneticum} simulation, mimicking a GAMA-like survey \citep{2022MNRAS.513..439D}. All algorithms are highly effective in identifying the majority of the group population down to halo masses of $M_{200}\sim10^{12.5}$ $M_{\odot}$, achieving completeness levels above 80\% with less than 20\% contamination from spurious objects. \cite{paola_stacking_magneticum} further demonstrate, using a mock dataset, that stacking optically selected groups in eROSITA data according to their halo mass accurately reconstructs the input X-ray surface brightness profile. In this study, we extend the stacking analysis to group spectra to measure the mean temperature of the hot gas. Previous spectral stacking attempts have been limited to the cluster-mass scale \citep{2014ApJ...789...13B}. We further extend the analysis to the group-mass scale, aiming to constrain the $M-T$ relation down to faint low-mass systems by capturing the bulk of the group population through the optical selection technique.

The organization of this paper is as follows. Sections \ref{sec:description_obs} and \ref{sec:description_sims} detail the observed and mock datasets employed in the analysis. Sections \ref{sec:method} and \ref{sec:tests} outline the spectral stacking technique and the tests conducted on the mock dataset. Section \ref{sec:results} presents our results for the $M-T$ relations based on the stacking analysis, discusses these results, and compares them with existing literature and predictions from state-of-the-art hydrodynamical simulations. Finally, Section \ref{sec:conclusions} summarizes our conclusions. Throughout this paper, we assume a flat $\Lambda$CDM cosmology with $\Omega_m = 0.27$ and $H_0 = 70\;\mathrm{km}\;\mathrm{s}^{-1}\;\mathrm{Mpc}^{-1}$. {All logarithms are presented in base 10.}
   
\section{The observed dataset}\label{sec:description_obs}

{In this study, we analyzed X-ray data for galaxy groups that were initially identified through optical selection. In the following section, we provide an overview of the optical and X-ray datasets used.}

\subsection{The optically selected group sample}
\label{optical}
We used the catalog of \cite{yang_galaxy_2007}, hereafter Y07, as the base for our observational sample. This catalog employs the halo-based group finder developed by \cite{2005MNRAS.356.1293Y} to select galaxy groups from the Sloan Digital Sky Survey (SDSS DR4; \citealt{sdss4}). It contains a total of 472,507 groups, including systems of isolated galaxies, and provides two proxies for the halo mass of the identified groups: one based on the characteristic stellar mass and the other on the characteristic luminosity. The total optical luminosity is provided in the R band, with an absolute magnitude cut at $M_R = -19.5$ mag. Stellar mass estimates were derived from the MPA-JHU catalog \citep{2004MNRAS.351.1151B}. The total luminosity and mass, obtained by summing the contributions of all members of the group, were used as proxies for the halo mass through the calibrations presented in \cite{2005MNRAS.356.1293Y}. To assess the robustness of the sample, \citet{2009ApJ...695..900Y} present luminosity and stellar mass functions of galaxies in groups, which are consistent with previous studies. In addition, we verified the performance of the group-finding algorithm using simulations, as discussed in Section~\ref{optical_mock}. In this work, we used the halo mass determined from the characteristic luminosity through the mass-to-light ratio.

From the Y07 catalog, we selected only sources in the eROSITA-DE sky coverage, with halo mass $M_{200} > 10^{13.0} M_{\odot}$ and a redshift limit of $z<0.2$. This mass limit ensures that temperatures can be effectively measured within the most sensitive energy range of eROSITA, while the redshift cut maximizes completeness over the halo mass range. These criteria yielded a total of 26,822 groups. 

Next, we recovered the size and mass of the identified groups. Following \cite{yang_galaxy_2007}, we estimated $R_{180}$\footnote{{Here, $R_{\Delta}$ is the radius within which the mean density of the halo is $\Delta$ times higher than the critical density of the Universe.}} using the formula:
\begin{equation}
R_{180} = 1.26\,h^{-1}\,\mathrm{Mpc} \left( \frac{{M_h}}{{10^{14}\,h^{-1}\,M_{\odot}}} \right)^{1/3}(1+z_{\mathrm{group}})^{-1},
\end{equation}

where $z_{\mathrm{group}}$ is the group redshift and $h = 0.73$ \citep{yang_galaxy_2007}. To convert $R_{180}$ to $R_{500}$, we followed the methodology outlined by \citet{2013SSRv..177..195R}.

\begin{figure}
   \centering
   \includegraphics[width=1\hsize]{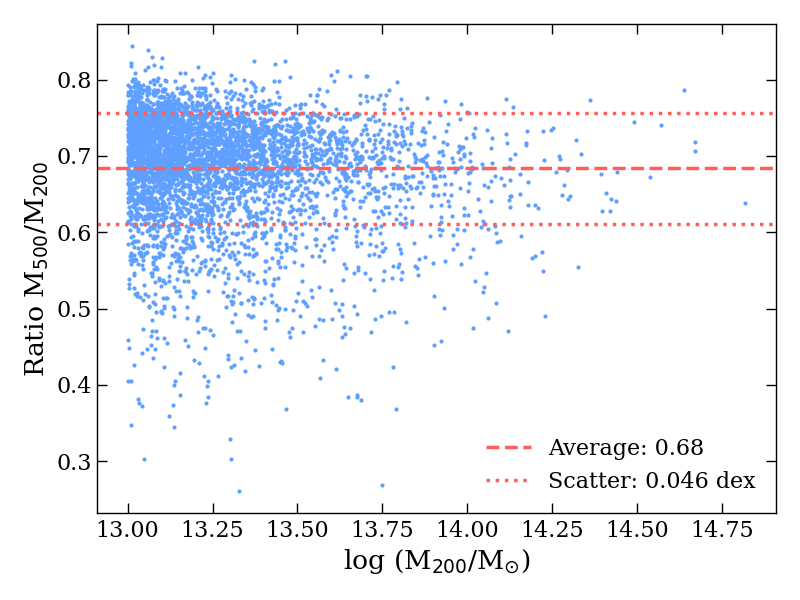}
      \caption{{Distribution of mass ratios for halo masses within $R_{500}$ and $R_{200}$ from the {\sc Magneticum} simulations. The horizontal red line indicates the average value used to convert $M_{200}$ to $M_{500}$ in this work. \label{fig:m200_to_m500}}}
   \end{figure}

   \begin{figure}
   \centering
   \includegraphics[width=1\hsize]{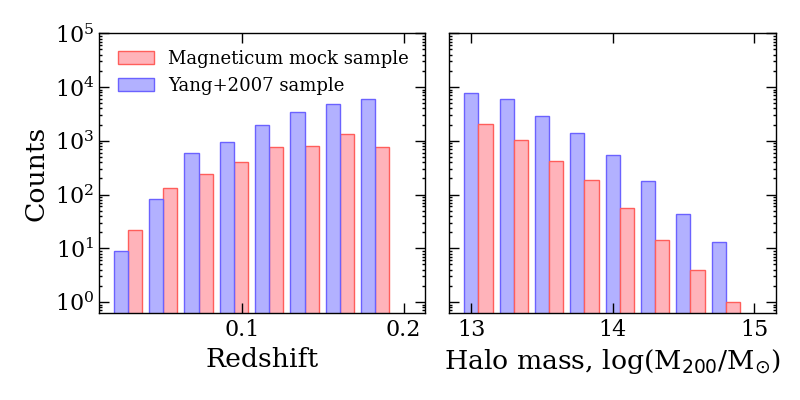}
      \caption{Distribution of sources used in this work with redshift and gas mass within $R_{200}$. The sample based on \cite{yang_galaxy_2007}, described in Section \ref{sec:description_obs}, is plotted in blue. The sample from the {\sc Magneticum} lightcone, described in Section \ref{sec:description_sims}, is shown in red. \label{sample}}
   \end{figure}

{The halo masses provided in the Y07 catalog are defined up to an overdensity of 180, whereas in this work we aimed to use the overdensity within $R_{500}$. Converting the mass estimated with $R_{180}$ into the mass within $R_{500}$ would require assuming a DM density profile. Because this study covers a wide range of halo mass, from low-mass groups to massive clusters, assuming a standard DM profile, such as the Navarro–Frenk–White (NFW) profile, would require knowledge of the halo concentration, which can vary both at fixed halo mass and across the mass range. To avoid making such assumptions, we used the {\sc Magneticum} hydrodynamical simulation to estimate a mass-dependent correction. We assumed that $M_{180}\sim M_{200}$ and we used the halo mass provided by the {\sc Magneticum} halo catalog (see Section \ref{sec:description_sims}), to estimate the ratio $M_{500}/M_{200}$. Figure \ref{fig:m200_to_m500} shows how this ratio varies with the halo mass $M_{200}$. The ratio remains $\sim0.7$ throughout the halo mass range covered in this study, with a scatter of 0.046 dex. The halo mass and redshift distribution obtained with this approach are shown in Figure \ref{sample}. 

{To validate our procedure, we compared the resulting $M_{500}$ and $R_{500}$ values with those obtained using the {\sc Python} package {\sc Colossus} \citep{2018ApJS..239...35D}, also based on $M_{180}$. The comparison showed a high level of agreement, indicating that our $M_{500}/M_{200}$ conversion is not strongly sensitive to the specifics of the simulation physics, such as feedback models.} 

\subsection{The eRASS1 data and catalogs}\label{sec:erass1_data}
The eROSITA All Sky Survey DR1 (eRASS1) covers an area of 20,626.5 square degrees, exhibiting highly variable effective exposure times ranging from approximately 100 seconds at the ecliptic equator to more than 10,000 seconds near the ecliptic pole within the 0.6--2.3 keV energy band \citep{2024A&A...685A.106B}. The eRASS1 data were processed with the standard eROSITA Science Analysis Software System \citep[eSASS,][]{2022A&A...661A...1B}. The data were sorted into overlapping sky tiles of size $3.6 \times 3.6$ deg$^2$, covering the energy range 0.2--10 keV, and providing calibrated event lists, along with images, exposure maps, background maps, sensitivity maps, and flux upper limits across various energy bands. 

{The eROSITA DR1 includes corresponding point source and extended emission catalogs {\citep{2024A&A...682A..34M, 2024A&A...685A.106B}}. Point source detection is based on the eSASS detection routine. The catalog contains $\sim$930,000 sources over the southern hemisphere. Adopting a uniform all-sky flux limit (at 50\% completeness) of F$_{0.5-2 \mathrm{keV}} > 5 \times 10^{-14}$ erg s$^{-1}$cm$^{-2}$, eROSITA resolves about 20\% of the cosmic X-ray background into individual sources in the 1--2 keV range. The group and cluster catalog is described in \cite{2024A&A...685A.106B} and \cite{2024A&A...688A.210K}. It contains $\sim$12,250 optically confirmed galaxy groups and clusters detected in the 0.2--2.3 keV band as extended X-ray sources in a 13,116 deg$^2$ region in the western Galactic half of the sky, which eROSITA surveyed in its first six months of operation. The clusters in the sample span the redshift range $0.003 < z < 1.32$. The mass range of the sample
covers three orders of magnitude, from a few very nearby local groups at $5 \times10^{12}$ $M_{\odot}$ to several massive clusters at $2 \times 10^{15}$ $M_{\odot}$ across a wide redshift range. }

\section{The mock dataset\label{sec:description_sims}}

\subsection{The L30 Magneticum lightcone}\label{sec:L30}
To validate each step of the optical group selection and stacking procedure in the eROSITA data, we generated a mock optical catalog and a mock eROSITA observation based on the same lightcone from the {\sc Magneticum} simulation, producing analogs of the observational dataset. Description of the lightcones and details of the mock observation design are given in \cite{ilaria_lightcone} and \cite{ilaria_opticallightcone}. Here, we provide a brief description of {\sc Magneticum} and a summary of the results and properties of the mock observations and catalogs. }

The {\sc Magneticum Pathfinder} simulation\footnote{\url{http://www.magneticum.org/index.html}} represents a state-of-the-art suite of cosmological hydrodynamical simulations performed using the P-GADGET3 code \citep[][]{springel_cosmological_2005}. Key advancements include a higher-order kernel function, as well as time-dependent artificial viscosity and conduction schemes \citep{2005MNRAS.364..753D}. These simulations incorporate a range of subgrid models to address unresolved baryonic physics, such as radiative cooling \citep{wiersma_effect_2009}, a uniform time-dependent UV background \citep{haardt_modelling_2001}, star formation and stellar feedback (e.g., galactic winds; \citealt{springel_cosmological_2003}), and explicit chemical enrichment from stellar evolution \citep{tornatore_chemical_2007}. Moreover, they feature models for the growth, accretion, and AGN feedback of supermassive black holes (SMBHs), following established approaches \citep{springel_cosmological_2005,di_matteo_energy_2005,fabjan_simulating_2010,hirschmann_cosmological_2014}. For a detailed description of the AGN feedback implementation, including black hole (BH) accretion, energy release mechanisms, and the transition from quasar-mode to radio-mode feedback, see \citet{2025arXiv250401061D}.

Here, we briefly describe the synthetic optical and X-ray datasets to emphasize their similarities with the observed data. The description refers to the L30 lightcone over an area of 30$\times$30 deg$^2$ up to $z=0.2$, thus simulating only the local Universe. Detailed information on the generation of the mock optically selected group catalog is provided in \cite{ilaria_opticallightcone}, while \cite{ilaria_lightcone} gives a detailed description of the mock X-ray observations derived from the same lightcone.

\subsection{The mock galaxy catalog and galaxy group sample}
\label{optical_mock}
The mock galaxy catalog was derived from the lightcones of the {\sc Magneticum} simulation. The galaxy and halo catalogs within {\sc Magneticum} were identified using the SubFind halo finder \citep{2001MNRAS.328..726S,2009MNRAS.399..497D}, which compiles a comprehensive list of observables (e.g., stellar mass, halo mass, and star formation) by integrating the properties of the bound particles. The mock galaxy catalog generated from the lightcone was limited to the local Universe up to $z < 0.2$ and covers an area of 30$\times$30 deg$^2$. It includes synthetic absolute rest-frame magnitudes in the SDSS filters (u, g, r, i, z), observed redshifts, stellar mass, and projected positions on the sky, i.e., right ascension (RA) and declination (Dec), for each galaxy. 
We limited the galaxy sample to stellar masses of $\geq 10^{9.8} M_{\odot}$ to ensure completeness for {\sc Magneticum}'s stellar mass resolution. To mimic observational uncertainties, we assigned errors to the observed redshifts and stellar masses, drawn from Gaussian distributions with $\sigma = 45$ km s$^{-1}$ and 0.2 dex, respectively. Additionally, 5\% of the galaxies were set to undergo catastrophic failure in the spectroscopic survey (i.e.,$\Delta v > 500$ km s$^{-1}$), and a spectroscopic completeness of 95\% was simulated.

To create an optically selected galaxy group catalog analogous to the Y07 galaxy group sample, we applied the same galaxy group finder algorithm as in \cite{2005MNRAS.356.1293Y} to the mock galaxy catalog described above. The algorithm's performance was thoroughly tested in \cite{ilaria_opticallightcone}  by matching the input halo catalog with the group catalog in terms of coordinates, redshift, and mass. Here, we report the main results regarding the algorithm's performance in terms of completeness, contamination, and the best halo mass proxy, as these are particularly relevant for the stacking analysis in the corresponding mock eROSITA observations.

\begin{figure}
   \centering
   \includegraphics[width=0.53\hsize]{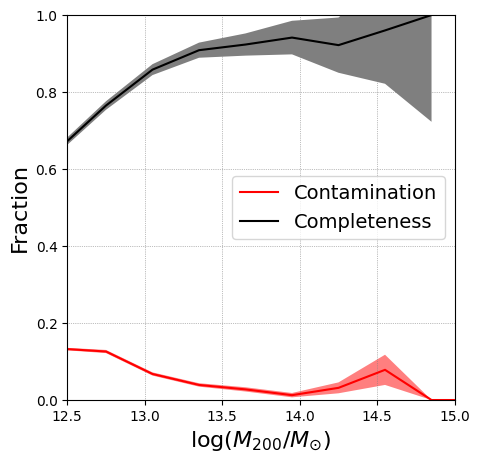}
    \includegraphics[width=0.46\hsize]{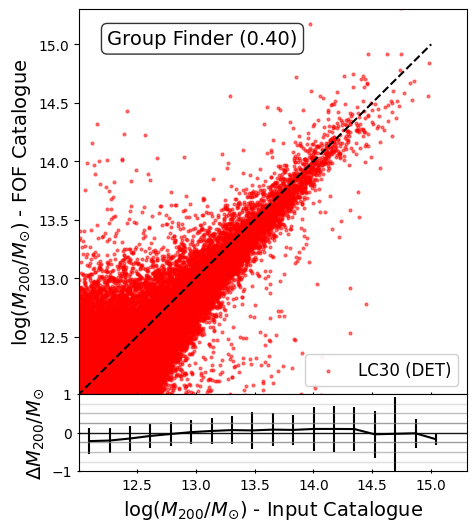}
      \caption{{Figures adapted from \cite{ilaria_opticallightcone}. \textit{Right panel:} Completeness (black) and contamination (red) as a function of halo mass for \cite{2005MNRAS.356.1293Y}'s Friends-Of-Friends (FoF) algorithm results applied to the {\sc Magneticum} mock optical catalog. The shaded region corresponds to the 95th binomial confidence interval. \textit{Left panel:} Comparison of the halo mass estimated by the \cite{2005MNRAS.356.1293Y} FoF algorithm as a function of the true input mass from {\sc Magneticum}. Masses are estimated from the total stellar luminosity proxy.} \label{fig:completeness}}
\end{figure}

As highlighted by \cite{ilaria_opticallightcone}, completeness remains above 75\% over the entire halo mass range probed, down to Milky Way group-sized halos with masses of $M_{200}\sim10^{12.5}$ $M_{\odot}$ (see Fig. \ref{fig:completeness}). Contamination due to spurious detections remains below 20\% to $M_{200} \sim 10^{13}$ $M_{\odot}$ and reaches 15\% at masses above $M_{200} \sim 10^{14}$ $M_{\odot}$. According to the analysis in \cite{ilaria_opticallightcone}, this is due to the algorithm fragmenting massive clusters in their substructures and misclassifying them as independent groups. In addition, according to \cite{ilaria_opticallightcone}, both halo mass proxies provided by \cite{yang_galaxy_2007}, based on total optical luminosity and stellar mass, show excellent agreement with the input halo mass, making the Y07 sample an ideal prior sample for stacking. 

\subsection{The synthetic eROSITA observation}
The synthetic eROSITA observation was generated by simulating the telescope's scanning strategy and observation mode of an ideal distribution of photons extracted from the X-ray-emitting components in the simulation. The exposure time was equal to an eRASS:4 observation (where the number four denotes the cumulative data from the four full-sky passages). The photon list of all X-ray emitting components, including hot gas, AGN, and X-ray binaries, was generated using {\sc PHOX} {\citep{2012MNRAS.420.3545B, 2013MNRAS.428.1395B, 2018MNRAS.481.2213B, 2023A&A...669A..34V}} for the same L30 lightcone used to create the mock galaxy catalog. The {\sc PHOX} software computes X-ray spectral emission based on the physical properties of the gas, BHs, and star particles in the simulation. Detailed modeling of the components can be found in \cite{ilaria_lightcone}.

The synthetic photon lists were used as input files for the Simulation of X-ray Telescopes ({\sc SIXTE}) software package v2.7.2 \citep{2019A&A...630A..66D}. {\sc SIXTE} incorporates all instrumental effects, including the point spread function (PSF), redistribution matrix file (RMF), and auxiliary response file (ARF) of the instrument \citep{2021A&A...647A...1P}. It can also model eROSITA's unvignetted background component due to high-energy particles \citep{2022A&A...661A...2L}. We performed mock observations of eRASS:4 in scanning mode using the theoretical attitude file for the three components separately, then combined the event files. In
L30, the simulated background in all seven telescope modules (TMs) is based on \cite{2022A&A...661A...5L} and represents the spectral emission from all unresolved sources, rescaled to the eRASS:4 depth in line with the simulated emission of the individual X-ray emitting components.

The simulated eROSITA data was processed through eSASS as described in \cite{2024A&A...682A..34M}. The event files from all emission components and eROSITA TMs were merged and filtered for photon energies within the $0.2-2.3$ keV band. The filtered events were binned into images with a pixel size of $4\arcsec$ and $3240 \times 3240$ pixels. These images correspond to overlapping sky tiles of size $3.6 \times 3.6$ deg$^{2}$, with a unique area of $3.0 \times 3.0$ deg$^{2}$. The L30 was covered by 122 standard eRASS sky tiles. A detailed description of the data reduction is provided in \cite{ilaria_lightcone}, along with the corresponding X-ray catalog of extended and point sources provided by eSASS.

\cite{ilaria_lightcone} thoroughly analyze the completeness and contamination of the extended emission catalog. After matching the X-ray detections with the input halo catalog, they find that the sample's completeness drops below 80\% at $M_{200} \sim 10^{14} M_{\odot}$, reaches 45\% at $\sim 10^{13.5} M_{\odot}$, and no sources are detected at $\sim 10^{13} M_{\odot}$. The contamination is negligible in the cluster mass range and reaches approximately 20\% for $10^{13} M_{\odot} < M_{200} < 10^{14} M_{\odot}$. This is in line with the results obtained by the eROSITA Consortium studies{\citep{2022A&A...665A..78S, 2018A&A...617A..92C, 2024A&A...691A.188B}}. Nevertheless, the eRASS1 data suffer from a more limited completeness and higher contamination due to the shallower nature of the observation. This is clearly highlighted in \cite{2024A&A...685A.106B}, which emphasizes that the completeness level of the eRASS DR1 group and cluster sample in the local Universe is limited to the cluster regime and comprises only a negligible fraction of the group population. For this reason, building an average view of the X-ray appearance and spectral properties of groups requires stacking systems selected through a different technique, as highlighted in \cite{paola_stacking_magneticum}.

\section{Prior sample for stacking \label{sec:filt}}

As discussed in \cite{paola_stacking_magneticum, paola_gasfraction, paola_profiles}, the X-ray emission of galaxy groups can be highly contaminated by other sources, in particular X-ray AGNs and X-ray binaries. To reduce such contamination, we selected only groups that satisfy the following criteria. Within projected $R_{500}$, the system does not contain:

\begin{itemize}
    \item any other groups or clusters from the same catalog. This criterion reduces contamination by neighboring sources. 
    \item any point sources detected in eRASS1. In both cases, point sources were defined as those with extension likelihood $\mathrm{EXT_{LIKE}}=0$ and detection likelihood $\mathrm{DET_{LIKE}}\geqslant6$. Point sources inside $R_{500}$ would cover a significant fraction of halos due to eROSITA's {PSF} size \citep{2024A&A...682A..34M,2022A&A...661A...1B}, particularly for central AGNs. By removing systems containing point sources, we minimized the risk of retaining photons only from the outskirts of the halo, which could lead to observational biases. 
\end{itemize}

This selection led to a final sample of 4,452 galaxy groups originating from the {\sc Magneticum} mock optically selected group sample and 18,808 galaxy groups from the Y07 catalog. Further details on the implications of this selection for the final results are detailed in Section \ref{sec:mt}.

\section{The X-ray spectral stacking procedure}\label{sec:method}

The eRASS1 data are too shallow to detect most of the group population in the local Universe at $z < 0.2$ and to achieve a sufficient signal-to-noise ratio (S/N) in the spectra of individual detected sources. Therefore, this study aimed to stack the spectra of the galaxy groups within the halo mass bins and to derive an average group spectrum for temperature measurement. Spectral stacking was performed independently of source detection in the eRASS1 catalog, using coordinates from the optical catalogs and adopting $R_{500}$ as the spectral extraction radius. Here, we describe the spectral stacking procedure and validate it using the mock dataset. {Specifically}, we applied our procedure to the mock observations {to assess the consistency} between the input and output spectra and {investigate the sources of uncertainty} in our derivation of the halo gas temperature.

\begin{figure}
   \centering
   \includegraphics[width=1\hsize]{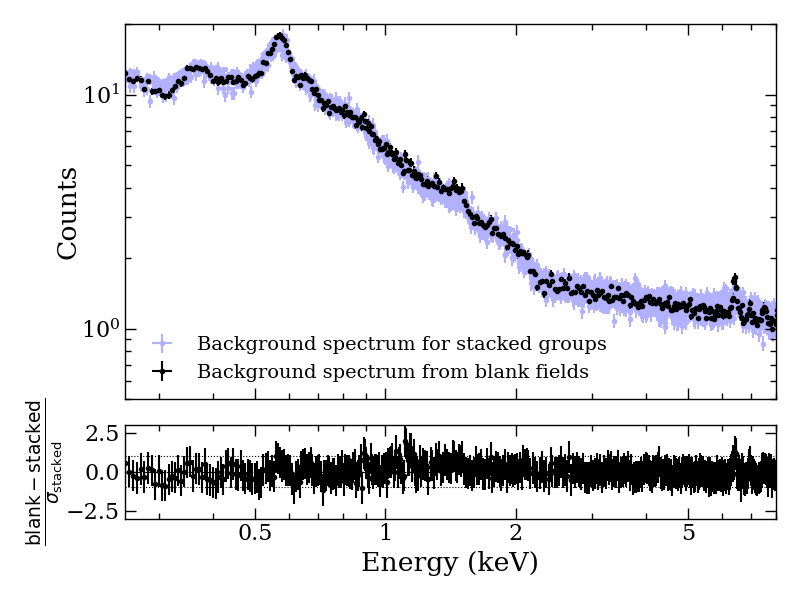}
      \caption{{\textit{Top panel:} Comparison of all background spectra used for different mass bins after the stacking procedure (violet points), along with the background spectrum from the stacking of 467 random blank eROSITA fields (black points). {\textit{Bottom panel}: Residuals, calculated as the difference between the blank fields spectrum and the stacked group spectrum, normalized by the uncertainty in the stacked spectrum.}} \label{fig:bkg_comparison}}
\end{figure}

\subsection{Description of the stacking procedure\label{sec:stacking}}

Before stacking, we cleaned all eRASS1 event lists of point sources by masking them with the fixed size of the average PSF \citep{2024A&A...682A..34M, 2022A&A...661A...1B} of eROSITA. This was carried out using the {\sc evtool} task of eSASS \citep{2022A&A...661A...1B} to remove point sources located outside $R_{500}$ of the selected sample. Point sources were selected from the eRASS1 point source catalog as described in Section \ref{sec:filt}. Masking was necessary to clean the background spectra of contamination from AGN and XRB photons. {For the mock sample, we performed masking based on point source detection in the X-ray mocks equivalent to eRASS:4.} 

From the masked event lists, we extracted the X-ray spectrum for each source using the {\sc srctool} task from eSASS. For the source spectrum extraction, we employed a circular region with a radius equal to $R_{500}$ at the given redshift. To extract the background spectrum, we used an annular region of $1.2 R_{500} < r < 1 \mathrm{Mpc}$ if $1.5 R_{500} < 1 \mathrm{Mpc}$, and $1.2 R_{500} < r < 1.7 R_{500}$ if $1.5 R_{500} > 1 \mathrm{Mpc}$. This second configuration was necessary to gather sufficient background photons for the higher-mass end of the sample, where $R_{500}$ is relatively large. In addition to the source and background spectra, {\sc srctool} generates effective RMF and ARF for each source. 

After extracting the spectra, we transformed them, along with the background spectra, RMF, and ARF files, into the observer's rest frame (z = 0) using the redshifts provided in the parent optical catalogs. {To achieve this, we shifted the spectra and response matrices to the spectral channels corresponding to their rest-frame energy values, applying interpolation to account for missing channels.}

Stacking of redshifted spectra was performed using the {\sc combine\_spectra} tool from the Chandra Interactive Analysis of Observations \citep[CIAO,][]{ciao}, which can be used for other X-ray facilities as long as the formats of spectra, RMF, and ARF files are homogenized to the required input. As a result, for each halo mass bin we produced an X-ray spectrum for further analysis. In addition to the source spectra, it also stacks the RMF and ARF files, and the background spectra. The stacked event lists on the target coordinates for each mass bin are detailed in Appendix \ref{ap1}.

{To verify the robustness of our background subtraction methodology {and ensure that the outskirts of the clusters were not inadvertently included during the extraction of background spectra}, we performed several tests. Figure \ref{fig:bkg_comparison} shows the comparison between the background spectrum measured around the sources, as described above and redshifted to the average redshift of the sample ($z=0.15$) to account for the redshift correction described earlier, and the background obtained by stacking randomly selected regions in the eRASS1 data {within the Y07 footprint, which corresponds to the SDSS DR7 footprint,} free of {detected} point sources or extended emission. Specifically, we selected a circular region with a radius of 1000 arcsec for each of the 467 eROSITA tiles in the Y07 catalog area and stacked them following the same procedure described earlier, {but without blueshifting the spectra}. The shape and normalization of the stacked spectrum were consistent (black spectrum in Fig. \ref{fig:bkg_comparison}) within the error bars with the stacked background spectra for each mass bin, which {validates} the results.} 
{This consistency does not contradict the known variations of the sky background between different areas, as the sky area used for the blank-sky backgrounds is the same as the sky area from Y07 used for spectral stacking. The blurring of the stacked source background spectra due to the redshift distribution of the sample lies within the error budget. The $1\sigma$ dispersion around the mean spectrum ranges from 0.68 counts in the $<0.5$ keV energy range to 0.09 counts in the $>2$ keV energy range.}

\begin{figure}
   \centering
   \includegraphics[width=1\hsize]{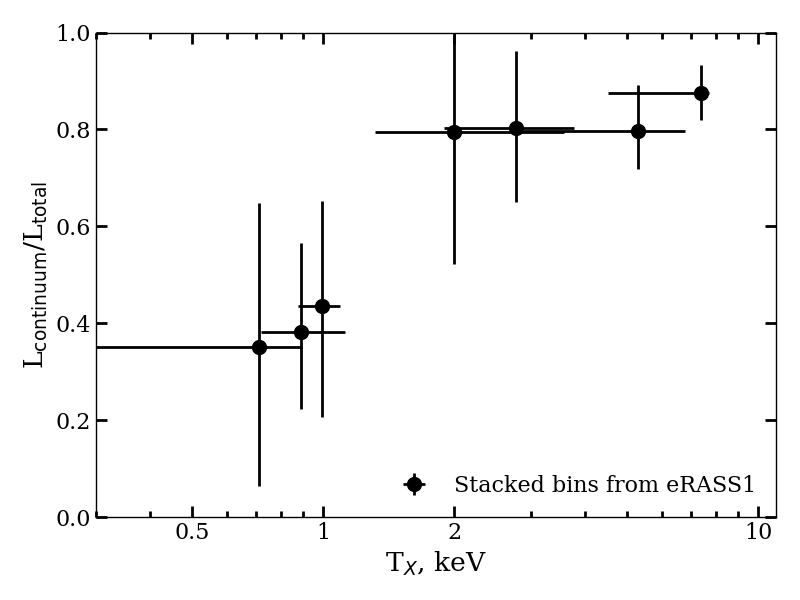}
      \caption{{Fraction of flux originating from the {continuum-only spectrum} relative to the total flux in the spectrum, as a function of {temperature}. The comparison shows that the contribution of emission lines to the spectrum is significant at low-temperatures. {The assumed metallicity is 0.3 solar.}} \label{fig:lines_fraction}}
\end{figure}

\subsection{Spectral modeling\label{sec:fitting}}

{ We adopted a twofold approach for our spectral {modeling} procedure: (i) {modeling} background-subtracted spectra, {where the model represents only the source itself}, and (ii) {modeling} the combined source plus background emission by fitting the final stacked spectrum as a sum of the source and a background spectrum blurred by the prior redshift distribution, as detailed in \cite{2014ApJ...789...13B}. Background modeling was expected to be more accurate, as it uses the combined spectrum of the source and the background,  and thus incorporates a larger number of photons into the analysis. {In addition, it avoids potential issues with error propagation that can arise in background subtraction, where the difference of two Poisson distributions can lead to non-trivial uncertainties.} Nevertheless, stacking a blueshifted source plus background spectrum at the source's rest frame causes a shifting of the background emission that does not originate at the same redshift. This implies that the final stacked spectrum is the result of the stacked rest-frame group emission plus a background component whose features are broadened by the redshift distribution, thereby complicating background spectrum modeling. 
We tested both approaches using the mock dataset and evaluated the consistencies and discrepancies of the results (see Section \ref{sec:tests}).

To estimate the temperature in each mass bin, we modeled the spectra using the Sherpa 4.16.0 Python package \citep{sherpa}. The stacked spectra were modeled using the {\sc gadem}\footnote\url{https://heasarc.gsfc.nasa.gov/xanadu/xspec/manual/XSmodelGadem.html} model, which represents a multi-temperature, collisionally ionized diffuse gas emission spectrum by combining multiple {\sc mekal}\footnote{\url{https://heasarc.gsfc.nasa.gov/docs/xanadu/xspec/manual/XSmodelMekal.html}} components \citep{1985A&AS...62..197M,1986A&AS...65..511M,1995ApJ...438L.115L} with a specified temperature distribution. We assumed a metal abundance of 0.3 solar using the abundance table from \cite{1989GeCoA..53..197A}.} This was combined with a multiplicative photoelectric absorption component, modeled using {\sc phabs}\footnote{\url{https://heasarc.gsfc.nasa.gov/docs/xanadu/xspec/manual/XSmodelPhabs.html}} with cross sections from \cite{1996ApJ...465..487V}. Further details on the metal abundances and absorption are provided later in this section. {In the {\sc gadem} model, the emission measure distribution followed a Gaussian profile, with the mean temperature, $T_{mean}$, and temperature standard deviation, $T_{\sigma}$, as key model parameters. We set the standard deviation of the temperature distribution to a fixed value of 0.2 keV, based on the distribution of mass-weighted temperatures from the simulations.} 

{The {\sc gadem} model accounts for both continuum bremsstrahlung emission and contributions from emission lines, which can play a significant role in the spectra of low-temperature IGrM {\citep{lovisari_review}}. For low-mass groups and those considered in this analysis, the contribution from emission lines is significant up to $\log(M_{500}/M_{\odot}) \approx 14.0$. Figure~\ref{fig:lines_fraction} shows the average fraction of flux originating from continuum emission relative to the total emission. {To estimate the continuum contribution, we calculated the total flux using the {\sc gadem} model with all parameters fixed to the best-fit values, except for the metal abundance, which was set to zero to remove the emission lines. As expected, the emission from low-mass groups is dominated by line emission (about 70\%) below {$T_X = 1$ keV, while for clusters the emission becomes increasingly dominated by the continuum, reaching 80–90\% above $T_X = 2$ keV, where it saturates.}}}

{Since all groups contaminated by {detected point sources} were excluded from the analysis, our stacked spectra did not suffer from emission due to bright AGNs. The contribution of the cosmic X-ray background due to unresolved AGNs} should have been accounted for by accurate subtraction of the background component. Nevertheless, the {primary} contribution of {unresolved} low-luminosity AGNs {likely originates from} the central or satellite galaxies of {the group} sample {and} may contribute to the source's X-ray emission, potentially contaminating the average emission in the stacked spectrum. After testing the level and spectral shape of the possible AGN component in the eROSITA mock observation,  we decided against including an AGN component in the model, since adding an extra component would introduce an additional source of uncertainty without a clear improvement in the temperature estimates (see Section \ref{sec:phox} for a detailed discussion).

To estimate the hydrogen column density along the line of sight for each source, we used the {\sc nh} tool from the {\sc FTOOLS}\footnote{http://heasarc.gsfc.nasa.gov/ftools} software \citep{ftools}. The derived values correspond to the 2D HI4PI map, a full-sky HI survey by the HI4PI collaboration \citep{extinction_map}. {To account for the molecular hydrogen component, we applied the relation between H$_2$ and HI from \cite{2013MNRAS.431..394W}. Additionally, we neglected the impact of self-absorption, as the sample is located at Galactic latitudes $l > 20^{\circ}$ \citep{extinction_map, 2005ApJ...626..195G}.}

\begin{figure}
\centering
\includegraphics[width=1\hsize]{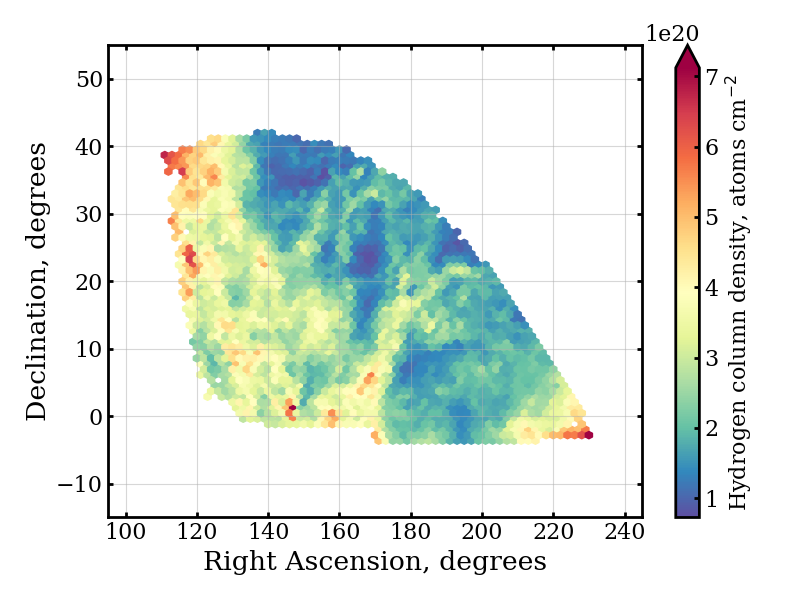}
\caption{Hydrogen column density map of all sources used in this study, based on the 2D HI4PI map. The color bar indicates the hydrogen column density for individual sources.
\label{fig:nh}}
\end{figure}

Our sample covers the SDSS region, which does not overlap with areas of high line-of-sight extinction, such as the Galactic plane and Galactic center. Therefore, all selected sources have column densities ranging from $1 \cdot 10^{20}$ to $8 \cdot 10^{20}$ atoms cm$^{-2}$, with an average of $2.6 \cdot 10^{20}$ atoms cm$^{-2}$. This indicates that {sources are weakly absorbed}. Figure \ref{fig:nh}. shows the column density map of the sample. For the fitting procedure, we used a fixed value of the hydrogen column density  {equal to an average of $2.6 \cdot 10^{20}$ atoms cm$^{-2}$}. To test the effect of local variations in column density, we also performed the same analysis using the minimum and maximum column density values, without observing any relevant difference.

\begin{figure}
\centering
\includegraphics[width=1\hsize]{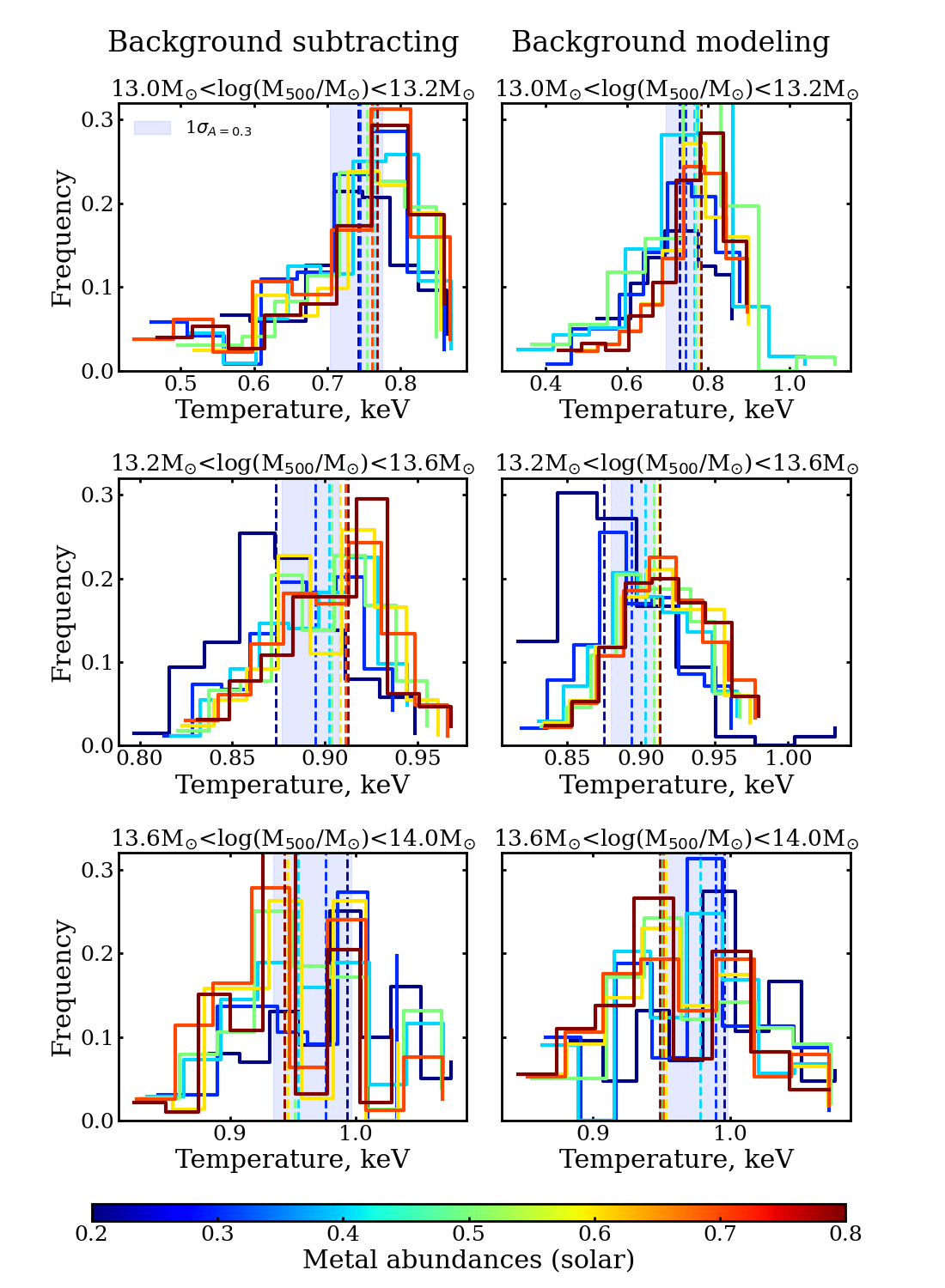}
\caption{Histograms of temperature values obtained using background subtraction (left column) and background modeling (right column) across different mass bins, showing the bootstrapped error distribution. The histograms are color-coded by metal abundances. Dashed vertical lines in corresponding colors indicate the mean temperatures. The blue shaded area highlights the 1$\sigma$ deviation for the abundance of 0.3 solar.\label{fig:abund}}
\end{figure}

In the fitting procedure, we neglected any metallicity radial profiles. We assumed a mean metallicity within $R_{500}$ of 0.3 solar, as recommended in \cite{abundances_1, gastadello_review, abundances_xcop, abundances_suzaku}. While the metal abundance varies significantly with radius, creating an abundance profile for the groups and clusters, its impact at $R_{500}$ is not significant. {Nevertheless, we tested the effect of this assumption and quantified its contribution to the error budget by comparing values of 0.2, 0.4, 0.5, 0.6, 0.7, and 0.8 solar with the 0.3 solar value (see Figure \ref{fig:abund}).}

To properly estimate the uncertainty without assuming any specific underlying error distribution, we used the bootstrapping method. For each mass bin, we performed spectral modeling on a stacked spectrum created by randomly sampling 68\% of systems in the original bin. From 200 independent bootstrapped samples, we took the median temperature as the final estimate, with the $3\sigma$ of the temperature distribution representing the error. This procedure, based on random sampling of 68\%, successfully reproduced the original standard deviations of the mass-weighted temperatures in the simulations. 

After estimating the temperatures, we applied a correction for systematic biases between the eROSITA and XMM-Newton data, as described in \cite{erosita_to_xmm}. This correction was necessary for comparing our $M-T$  results with the existing literature based on XMM-Newton and Chandra observations \citep{lovisari_review}, which are not significantly biased relative to each other in the low-temperature regime.

\section{Testing of the stacking procedure \label{sec:tests}}

Before applying the stacking procedure to the eRASS1 data, we validated it using the mock dataset. Specifically, we replicated all steps of our approach by stacking the mock optical group sample, based on the Y07 algorithm, onto the mock eROSITA observations and assessing the consistency between input and output. 

\subsection{Consistency of input and stacked spectra\label{sec:phox}}

\begin{figure*}
   \centering
   \includegraphics[width=0.49\hsize]{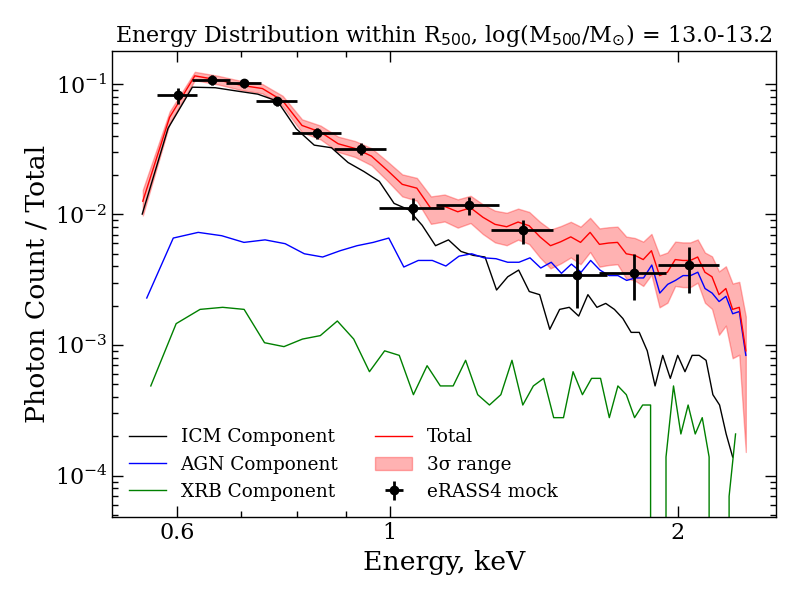}
   \includegraphics[width=0.49\hsize]{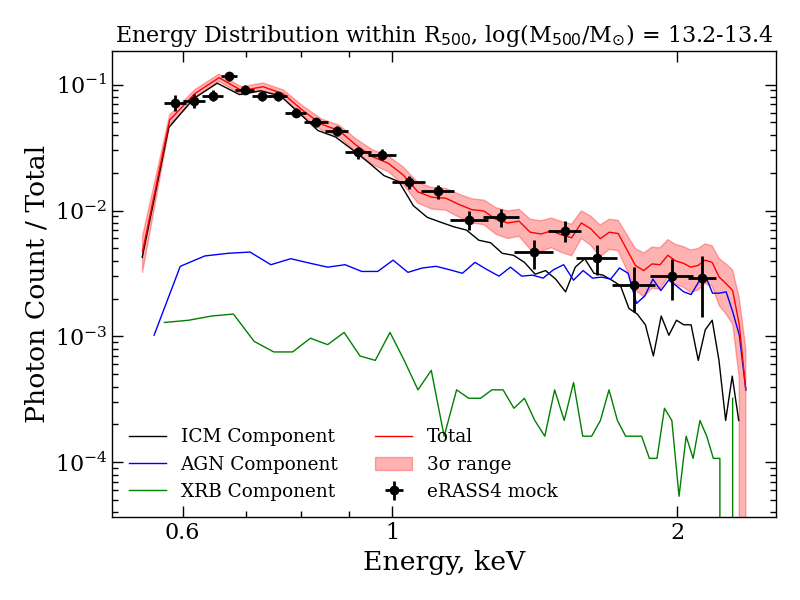}
   \includegraphics[width=0.49\hsize]
   {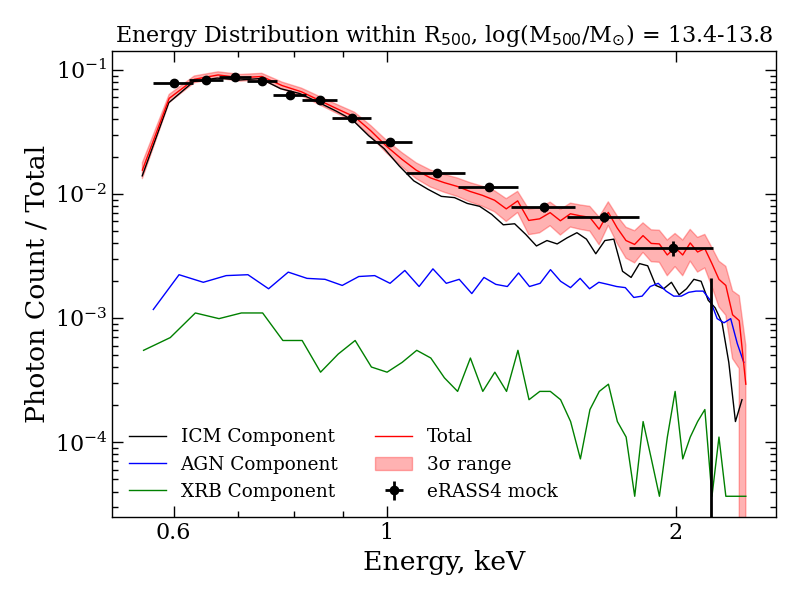}
   \includegraphics[width=0.49\hsize]{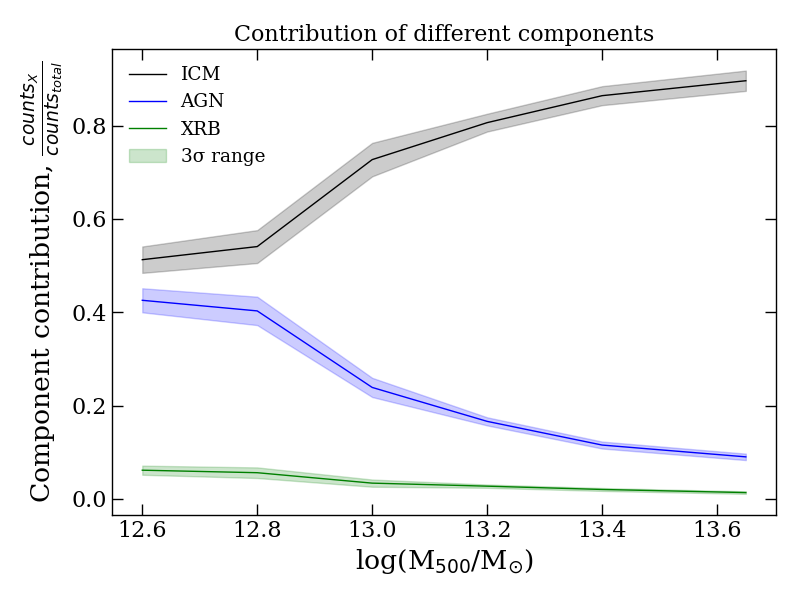}
      \caption{{{\it{Top panels and bottom left panel}}: Comparison of the PHOX spectrum with the resulting stacked spectrum for the mock sample across different mass bins. The red line represents the total PHOX X-ray spectrum, along with its 3$\sigma$ error range. The black, blue, and green lines correspond to the average PHOX X-ray spectra for the ICM, AGN, and X-ray binary (XRB) components, respectively. Black data points indicate the stacked mock X-ray spectrum, equivalent to eRASS:4 observations, after accounting for the eROSITA response curve. {The vertical axis corresponds to the fraction of photons within each energy bin from each component relative to the total counts in the 0.5--2 keV energy band.} {\it{Bottom right panel}}: Contribution of different spectral components to the total spectrum in the 0.5–2 keV range, shown after masking of point sources from the eRASS1 catalog. The black, blue, and green lines represent the IGrM, AGN, and XRB components, respectively.} 
      } \label{fig:phox_vs_mock2}
\end{figure*}

{First, we present the input spectra predicted by the {\sc Magneticum} simulation, generated using the PHOX tool for the lightcone described in Section \ref{sec:L30}, corresponding to the mass bins used in our analysis. The spectra shown in Fig. \ref{fig:phox_vs_mock2} are based on $\sim$2000 randomly selected halos from the {\sc Magneticum} simulation (Box2b/hr) with masses $M_{500}$ ranging from $10^{12.5}$ to $10^{14.5}$ $M_{\odot}$. The PHOX photon simulator also provides classifications for the origin of each photon, indicating whether they were emitted by the IGrM, AGN, or XRB components. This classification allows us to evaluate the contamination from each component in various mass bins, as well as their energy-dependent behavior.} {The {\sc Magneticum} simulation successfully reproduces observed AGN luminosity function \citep{2018MNRAS.481.2213B, hirschmann_cosmological_2014}.} {Figure \ref{fig:phox_vs_mock2} illustrates these input spectra, distinguishing the individual components. The AGN power-law contribution remains relatively flat in the soft energy band at 0.5--2 keV, while the XRB contribution is nearly negligible in the halo mass range considered here.}

{To validate the stacking procedure, we compared the resulting stacked spectrum for each mass bin in the {mock observation described in Section \ref{sec:description_sims}} with the corresponding input spectrum produced by PHOX, representing the original X-ray photons before accounting for the eROSITA response. The top panels and the bottom left panel of Fig. \ref{fig:phox_vs_mock2} show this comparison. {The stacked spectra (black points in the figure) were corrected for the eROSITA response.} The stacked spectra perfectly overlap with the uncertainty region of the averaged PHOX spectra, demonstrating that the stacking technique can retrieve the input spectra.}

{Using the information on the origin of photons (IGrM, AGN, or XRB), we assessed contamination levels across different mass bins. Figure \ref{fig:phox_vs_mock2} shows the fractional contributions of IGrM, AGN, and XRB photons to the total spectrum. Contamination from XRBs is negligible across all mass ranges. In contrast, AGN contamination is significant in the mass range below M$_{500}$ = $10^{13}M_{\odot}$. For halo masses exceeding $10^{13}M_{\odot}$, the AGN contamination decreases to 15-20\%.}. 

{The top panels of Fig. \ref{fig:phox_vs_mock2} show that, in the lower mass bin, AGN contamination becomes {more} prominent in the energy range above 1 keV. However, at lower energies, {its contribution is relatively smaller compared to the cluster emission}, particularly in the region of the iron line {complex}, which is important for the temperature measurements. In the higher mass bins, AGN contamination remains {relatively minor} across the entire 0.5–2 keV range \citep[see also][]{paola_stacking_magneticum}.}

{We note that all these contributions are calculated after applying the X-ray point source masking procedure described in Section \ref{sec:stacking} and for a sample cleaned of sources containing X-ray point sources within their $R_{500}$ (see Section \ref{sec:filt}). Consequently, the primary source of contamination identified here is the unresolved AGN population. It should also be noted that this mock analysis is based on eRASS:4-equivalent mock observations, implying that contamination levels are expected to be higher in the eRASS1 data.}

\begin{figure}
\centering
\includegraphics[width=1\hsize]{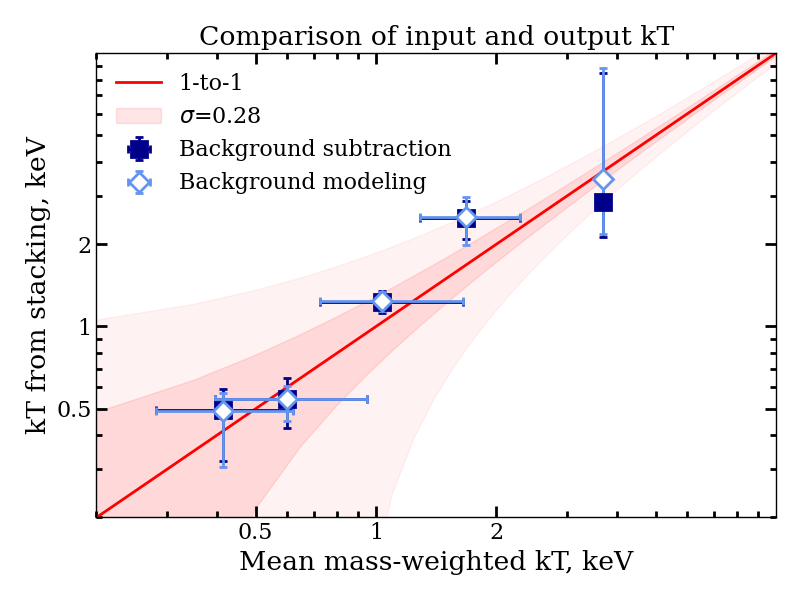}
\includegraphics[width=1\hsize]{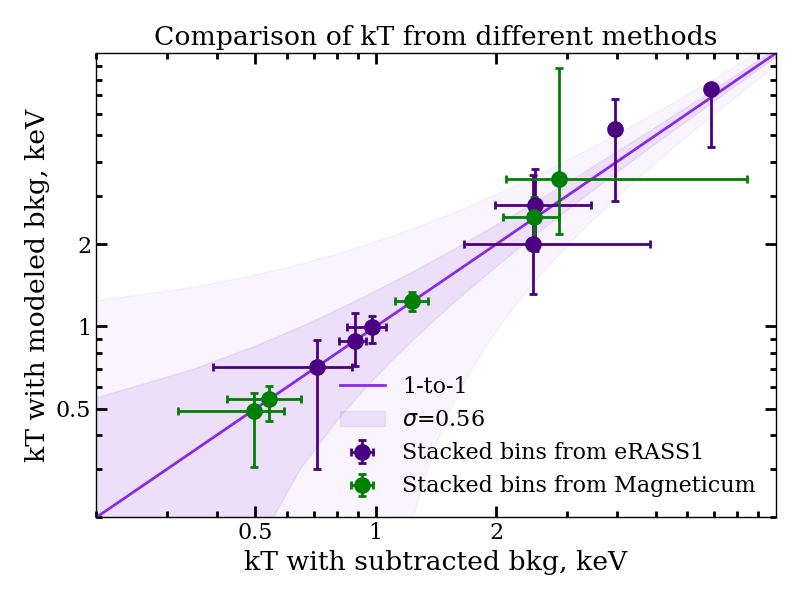}\\
\caption{{{\it{Upper panel}}: Comparison of temperatures derived from stacking with mean mass-weighted temperatures from simulations for the mock {\sc Magneticum} sample. Blue points show results from background subtraction, and black points represent results from background modeling. The red line shows the 1-to-1 relation, and the red shaded areas indicate the 1$\sigma$ and 3$\sigma$ error regions. {\it{Lower panel}}: Comparison of temperatures obtained for the observed {sample} ({violet} points) using background modeling and background subtraction methods. The red line represents the 1-to-1 relation, while the red shaded areas indicate the 1$\sigma$ and 3$\sigma$ error regions.} \label{fig:kT_kT}}
\end{figure}

\subsection{Testing of the temperature measurement}

{In this section, we test the spectral modeling and gas temperature measurements. We evaluate both approaches: one based on subtracting the background spectrum and another that models the background and source emission simultaneously. Previous studies have shown that the temperatures derived from individual spectra of massive clusters in the {\sc Magneticum} simulation are consistent with expectations \citep{2023A&A...675A.150Z}. This consistency confirms the reliability of individual spectra, allowing us to attribute any discrepancies in our tests to the spectral stacking process rather than the accuracy of the individual spectra themselves.}

First, for the simulated sources from {\sc Magneticum}, we compared the input mass-weighted temperatures with the temperatures derived from the stacked spectra. Mass-weighted temperatures were used instead of spectroscopic-like temperatures because the latter were not calibrated for the lower mass range of galaxy groups, although they perform well in the galaxy cluster mass range \citep{2004MNRAS.354...10M}. In the upper panel of Fig. \ref{fig:kT_kT}, we show this comparison for both the background-subtracted and background-modeled methods, including their corresponding 1$\sigma$ and 3$\sigma$ errors. In both cases, the measured temperatures are consistent with the input mass-weighted temperatures within the error bars. 

{We further verified the consistency between the results obtained using background modeling and background subtraction. In the lower panel of Fig. \ref{fig:kT_kT}, we compare the temperatures derived from the two methods for both the observed eRASS1 data and the simulated mock data. The results demonstrate consistency within the error bars for both datasets.}

\subsection{{Impact of possible spurious detections}}

{As described in Section \ref{optical_mock}, we anticipate a contamination level of approximately 10\%--15\% from spurious optical detections, depending on the mass bin, including fragmented halos and false-positive sources. To evaluate the impact of this contamination on temperature estimation, we conducted tests using simulated data. Specifically, we examined how the fraction of false-positive sources influences the resulting temperature measurements.}

\begin{figure}
\centering
\includegraphics[width=1\hsize]{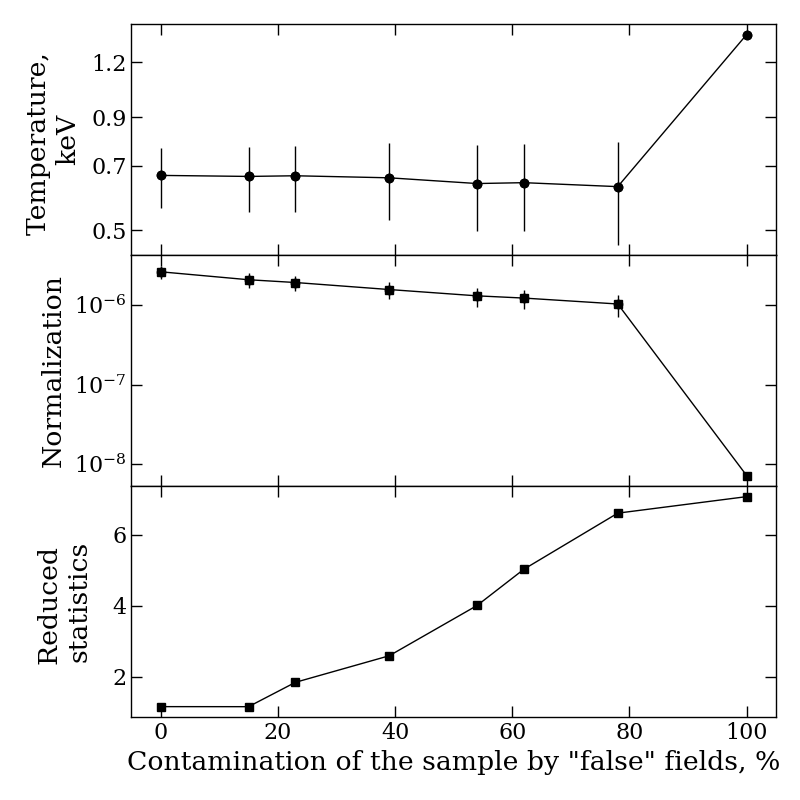}\\
\caption{{ Impact of contamination from spurious (``false'') fields on temperature, normalization, and fit statistics in the stacked sample from the {\sc Magneticum} mock observations equivalent to eRASS:4 depth. The top panel shows the measured temperature as a function of the fraction of false sources included in the sample. The middle panel presents the corresponding normalization values, which decrease with increasing contamination. The bottom panel shows the reduced statistic of the spectral fit, which increases with contamination. Error bars indicate 3$\sigma$ uncertainties.}
\label{fig:false_fields}}
\end{figure}

{For this test, we focused on the lower mass bin, stacking all spectra of sources used for temperature estimation from the mock observations alongside randomly selected blank fields. {We tested various levels of contamination, ranging from 10\% to 100\% (see Fig. \ref{fig:completeness}), where 100\% contamination means that the sample contains only blank field spectra. For all contamination levels, we} estimated the resulting temperatures, normalizations, and reduced statistics. As illustrated in Fig. \ref{fig:false_fields}, the inclusion of false sources does not significantly affect the measured temperatures.} At the same time, the presence of false sources increases the uncertainty by lowering the S/N of the stacked spectrum and increasing the reduced statistic of the spectral fit. As expected, the normlization of the spectrum decreases with increasing contamination from blank fields, indicating that, while contamination does not systematically bias temperature measurements, it does affect luminosity estimates. However, when attempting to fit a spectrum composed entirely of false detections (i.e., 100\% contamination) using the same initial model assumptions, the fit captures only noise, resulting in an unphysical temperature estimate, a normalization close to zero, and an inability to constrain the errors associated with temperature.
This confirms that the expected fraction of spurious sources does not affect the robustness of our method but may introduce additional uncertainty in error measurement through bootstrapping.

\section{Results and discussion}\label{sec:results}

{Here, we present the main $M-T$  relation from low-mass groups up to the cluster scale, using the temperature measurements estimated from the stacked spectra based on the Y07 galaxy group sample at $z < 0.2$. The derived temperatures in each mass bin are listed in Table \ref{tab:values}.}

\subsection{Mass-temperature relation across two decades of halo mass}\label{sec:mt}

\begin{figure}
   \centering
   \includegraphics[width=1\hsize]{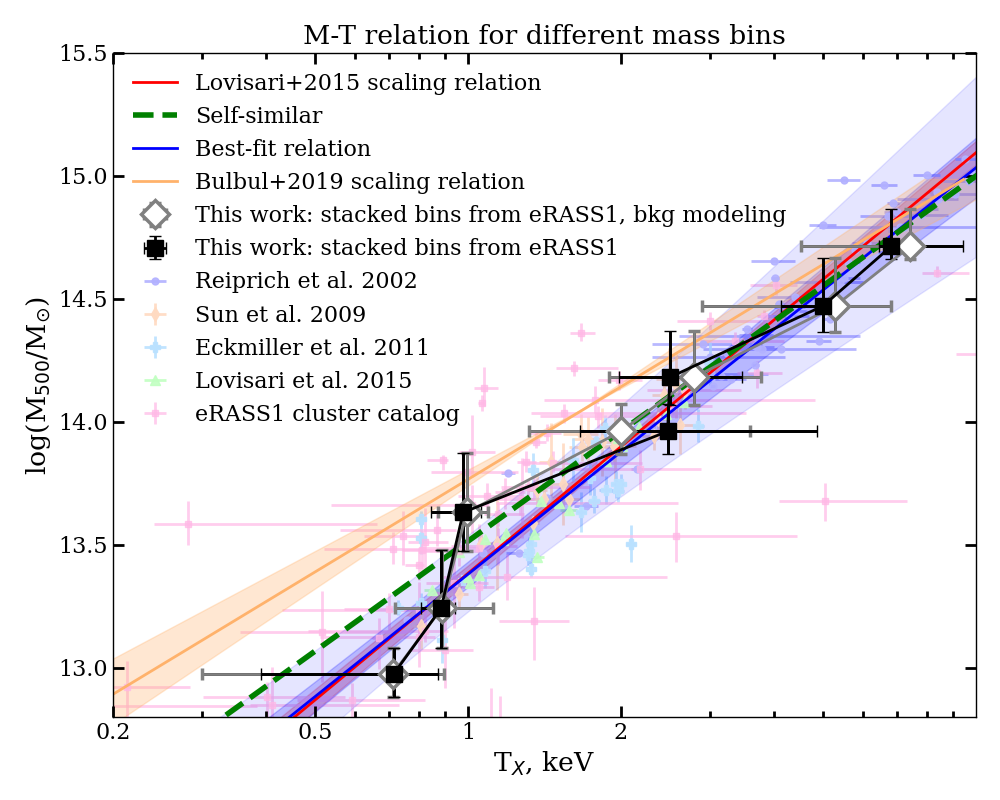}
      \caption{The mass-temperature relation for galaxy groups and clusters in observations, with mass defined inside $R_{500}$ as a function of X-ray temperature. Black {squares} correspond to stacked bins from eRASS1 observations. Open {diamonds} represent background modeling instead of background subtraction. {The red line shows the scaling relation from \citep{lovisari_relation}, the orange line represents the relation from \citep{2019ApJ...871...50B}, and the green dashed line indicates the self-similar model. The blue line corresponds to the best-fit relation. Shaded regions indicate the 1$\sigma$ and 3$\sigma$ confidence intervals for the respective relations.}\label{fig:m_t}}
   \end{figure}

{Fig. \ref{fig:m_t} shows the $M-T$ relation from the group mass scale with $M_{500} > 10^{13}\, M_{\odot}$ to the cluster regime at $10^{15}$  $M_{\odot}$. The figure shows the results for the temperature based on both stacking methods used here. The halo mass is based on the halo mass proxy of Y07 converted to $M_{500}$, as explained in Section \ref{optical}. The figure also includes a compilation of literature results, including temperature measurements of X-ray-selected clusters and groups. However, it is important to note that this compilation does not represent a homogeneous sample, as the sources were selected using different criteria, {and the masses and temperatures were derived using various techniques, such as optical analysis, gas density profiles, or weak lensing for mass estimates.} The compilation does include the most recent eRASS1 groups and clusters of \citet{2024A&A...685A.106B}. However, given the shallow nature of eRASS1 and the low S/N of the detected groups, we used only a subsample of extended sources with high S/N at $z<0.05$ with clearly defined optical counterparts. }

{{We fit the \textit{M–T} relation using Orthogonal Distance Regression (ODR) implemented via the {\sc 
 scipy.odr} module from the {\sc scipy} {\sc Python} package, which fully accounts for uncertainties in both temperature and mass as $x$ and $y$ axes.} The model assumes a log-linear relation $\log(y) = m \times \log(x) + b$. The best-fit relation is}

{
\begin{equation}
\log_{10} \left(\frac{M_{500}}{M_{\odot}}\right) = {1.65\pm0.11}\times \log_{10} \left( \frac{T_{X}}{1\;\mathrm{keV}} \right) + {13.38\pm0.05},
\label{eq:MT_relation}
\end{equation}}

{with an intrinsic scatter of $0.13{\pm}0.03$ dex, where all uncertainties are quoted at the 1$\sigma$ level. To verify the robustness of our results, we also performed a Bayesian linear regression using the Markov Chain Monte Carlo (MCMC) method implemented in the {\sc emcee} {\sc Python} package. The resulting slope, intercept, and intrinsic scatter are fully consistent with those obtained from the least-squares fit within the uncertainties, confirming the stability of our derived relation. {The relation and its confidence interval are shown in Fig.~\ref{fig:m_t} (blue line with shaded regions indicating the 1$\sigma$ and 3$\sigma$ intervals).}}

{The figure also shows the best-fit power-law relation of \citet{lovisari_relation} (red line) and the self-similar prediction (dashed green line). {The relation obtained in this work from the stacked spectra agrees within less than 1$\sigma$ across the two orders of magnitude in halo masses with the \citet{lovisari_relation} best-fit based on X-ray selected groups.} Thus, a single power law provides a good fits across the entire halo and temperature mass range sampled in this work. We note that, statistically, the observed data do not rule out {self-similarity}due to the relatively large error bars -- {the relation is in agreement with self-similar predictions within less than 2$\sigma$}.}

\begin{figure}
\label{sims}
   \centering
   \includegraphics[width=1\hsize]{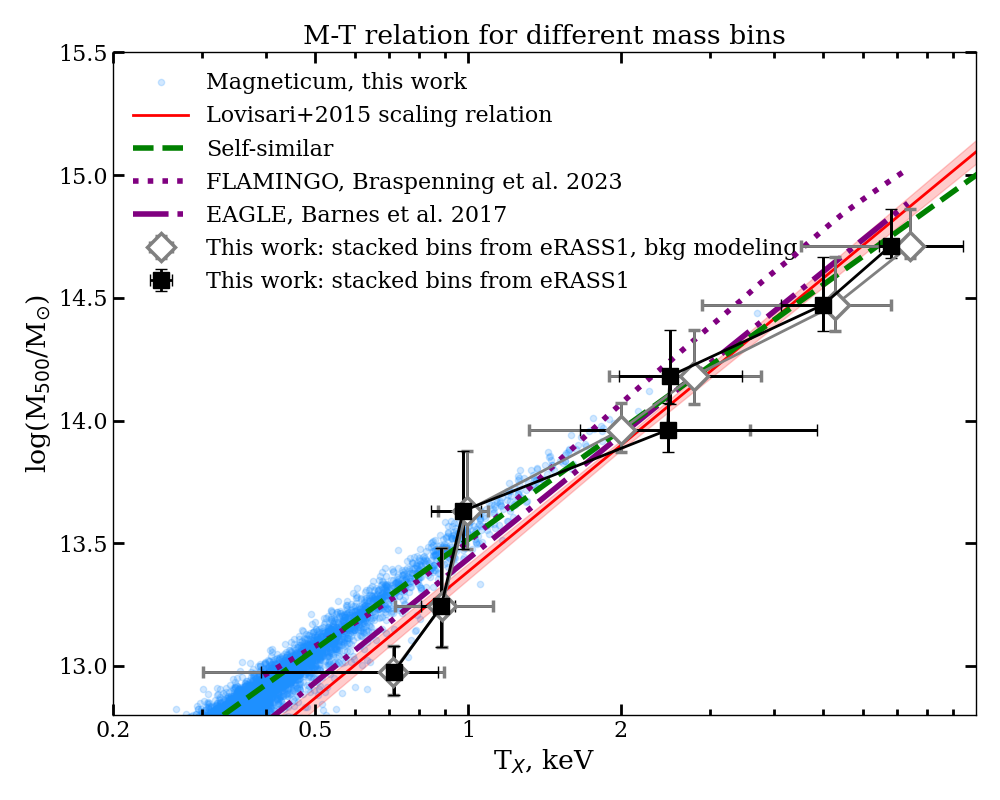}
      \caption{The mass-temperature relation for galaxy groups and clusters in simulations, with mass defined inside $R_{500}$ as a function of X-ray temperature. Blue background points represent individual mass-weighted temperatures for each simulated group in {\sc Magneticum}. Black {squares} correspond to stacked bins from eRASS1 observations. Open {diamonds} represent background modeling instead of background subtraction. The red line shows the scaling relation from \citep{lovisari_relation}, the green dashed line represents the self-similar model, and other lines depict results from FLAMINGO \citep{flamingo} and EAGLE \citep{eagle}. 
      \label{fig:m_t_sim}}
   \end{figure}

{It is unsurprising that the $M-T$ relation does not rule out the self-similar prediction. The $M-T$ relation mainly probes the temperature resulting from gravitational collapse. While nongravitational processes such as AGN feedback mainly affect the X-ray appearance of galaxy groups by {contaminating} the IGrM with metals or by expelling gas from the virial region \citep[see for instance][]{paola_gasfraction}, they may not be as effective in changing the overall halo gas temperature throughout the entire virial region of the system. The effect is likely limited to redistributing {baryons} within the system, displacing part of the baryonic matter beyond the virial radius, or affecting the temperature over a smaller region, such as the circumgalactic region of the central galaxy. Testing this hypothesis requires testing the X-ray luminosity -- temperature relation and constructing temperature and entropy profiles of systems at different mass scales. \cite{2024A&A...691A.188B} provide a first attempt at measuring the temperature and entropy profiles of eRASS1-detected groups with sufficient S/N. Nevertheless, \citet{ilaria_lightcone} {and \citet{2025arXiv250319121M}} point out that the X-ray selection might favor low-central-entropy systems at fixed halo mass, thereby inducing a potential bias in assessing the role of AGN feedback {and environment} in affecting the thermodynamical properties of the gas. Obtaining such profiles through stacking the bulk of the halo population could provide an unbiased view of the average effect of any nongravitational process on the gas temperature and density distribution.}

\subsection{{Insights from simulations on the $M-T$ relation}}

\begin{figure}
   \centering
   \includegraphics[width=1\hsize]{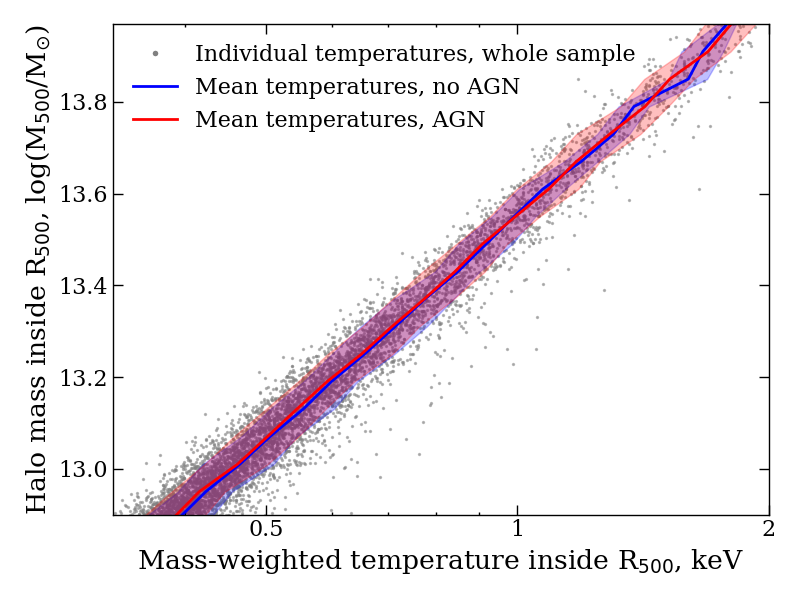}
      \caption{Comparison of the mean mass-weighted temperature as a function of halo masses for a sample of groups with and without X-ray AGN sources within their $R_{500}$. Grey points represent the individual mass-weighted temperatures for each source. The blue line indicates the mean temperature for the group without point sources from the mock catalog within their $R_{500}$, while the red line shows the mean for groups containing such sources. Shaded areas show the standard deviation of the mean values. All sources, along with their masses and temperatures, are derived from the {\sc Magneticum} lightcone.\label{agn_noagn}}
   \end{figure}

We note that in the current analysis we cannot probe the instantaneous effect of AGN activity on the gas temperature. Indeed, excluding systems hosting any point source prevents us from assessing whether AGN activity might have a direct observable effect. Nevertheless, we can test this hypothesis in the mock dataset, where AGN and IGrM emission can be disentangled, {and} examine the predictions of the {\sc Magneticum} simulation. Fig. \ref{agn_noagn} shows the {\sc Magneticum} $M-T$ relation for systems with and without current strong AGN activity in the central or satellite galaxies. By ``strong AGN activity,'' we refer to the presence of an X-ray point source within R$_{500}$, where point sources are defined as those with an extension likelihood of $\mathrm{EXT_{LIKE}}=0$ and a detection likelihood $\mathrm{DET_{LIKE}}\geqslant6$, as described in Section \ref{sec:filt}. We do not observe any indication that the {relation} is affected by ongoing AGN activity. This may suggest that the response time of the gas to AGN feedback is longer than the AGN duty cycle, or that the current activity is insufficient to affect the gas. Indeed, this is not surprising, as the accretion rate and feedback power in the local Universe are expected to be marginal compared to the high-redshift Universe. Thus, one would expect a stronger instantaneous effect at high redshift and a marginal impact in the local Universe.

It should be noted that the {\sc Magneticum} simulation performs well in reproducing various observed properties of AGN activity. As mentioned in Section \ref{sec:phox}, it accurately reproduces the observed AGN X-ray luminosity function up to z=0.5 \citep{2018MNRAS.481.2213B}. It also matches the BH mass function for log(M$_{\bullet}$/M$_{\odot}$)>7.5 at z<1 \citep{hirschmann_cosmological_2014}. Furthermore, \citet{paola_gasfraction} demonstrates that {\sc Magneticum} successfully reproduces the observed gas mass fraction across a wide range of halo masses, indicating that the strength and impact of AGN feedback are realistically modeled. These results suggest that {\sc Magneticum} mock observations provide a reliable framework for studying AGN activity and its effects and can be confidently applied to interpret the eRASS1 sample.

Fig. \ref{fig:m_t_sim} shows {a comparison between the data obtained in this work and the predictions from {\sc Magneticum}, as well as from other} hydrodynamical simulations: Full-hydro Large-scale structure simulations with All-sky Mapping for the Interpretation of Next Generation Observations (FLAMINGO) \citep{flamingo} and Evolution and Assembly of GaLaxies and their Environments (EAGLE) \citep{eagle}. 
{In both {\sc Magneticum} and EAGLE, energy is deposited thermally into particles or cells within the BH's sphere of influence. {\sc Magneticum} implements a two-mode AGN feedback model, with a quasar mode (high Eddington ratio) and a radio mode (low Eddington ratio) feedback \citep{fabjan_simulating_2010, 2025arXiv250401061D}, each with distinct feedback efficiencies. In contrast, EAGLE uses a single-mode feedback prescription. Meanwhile, FLAMINGO switches between thermal and kinetic feedback energy injection, depending on the accretion rate. More details on the comparison of different feedback implementations in simulations are provided by \citet{2025arXiv250206954V}.}

{Across the entire mass range, the {\sc Magneticum} results are consistent with the$M-T$ relation expected from self-similar predictions. The FLAMINGO predictions follow the self-similar slope at the low-mass end but tends to predict slightly lower temperatures for halos with M${500} > 10^{14}$ M${\odot}$. In contrast, the EAGLE predictions closely follow the observational relation from \citet{lovisari_relation} across the full mass range.}

\section{Conclusions}\label{sec:conclusions}

Using the eRASS1 dataset obtained during the first six months of the survey, we statistically explored the {low-mass} end of the $M-T$ relation. Using the {\sc Magneticum} simulation, we validated our stacking methods. {We fit our data to derive the  best-fit log-linear slope, intercept, and intrinsic scatter, as shown in Eq.~\eqref{eq:MT_relation}. For each mass bin, the results are close to the expected self-similar predictions without heating from feedback: the agreement between our best-fit relation and the theoretical self-similar predictions is 1.7$\sigma$ in the 0.2–10 keV energy range. In the same energy range, our results are statistically consistent with the relation from \cite{lovisari_relation} within 1.15$\sigma$, which is based solely on X-ray-detected sources.}

{We conclude that the combination of stacking, existing wide-field X-ray surveys, and optical-based sample selection enables the study of the X-ray properties of sources that are typically undetected.} This approach can be applied in future studies to both galaxy groups and other types of X-ray ``hidden'' objects, such as low-luminosity AGNs selected by optical spectra or high-redshift cluster candidates expected from the upcoming  Legacy Survey of Space and Time (LSST) and Euclid datasets.

In the long term,  eRASS:4 is expected to significantly increase the statistics in each mass bin. This should allow us to probe temperature profiles to study feedback more precisely, rather than relying solely on the average temperature inside the entire IGrM. Similar improvements are expected from upcoming spectroscopic facilities such as The Dark Energy Spectroscopic Instrument (DESI) and 4-metre Multi-Object Spectroscopic Telescope (4MOST), which will provide future extended galaxy group catalogs.

We do not detect a significant change in the $M-T$ relation slope; therefore, temperature can serve as a mass proxy across the entire mass range. This validates the use of temperature-derived masses in cosmological studies, substantially widening the mass range and improving cosmological parameter estimates. For example, for $S_8$, whose late-Universe measurements were found to be in tension with CMB-based studies \citep{s8_cosmic_shear}, recent eRASS1 estimates show agreement with the CMB \citep{erass1_cluster_counts}.

\begin{acknowledgements}
       VT and IM acknowledge support from the European Research Council (ERC) under the European Union’s Horizon Europe research and innovation programme ERC CoG (Grant agreement No. 101045437, PI P. Popesso). VT acknowledges Kirill Grishin for the useful discussions and technical support. KD acknowledges support by the COMPLEX project from the ERC under the European Union’s Horizon 2020 research and innovation program grant agreement ERC-2019-AdG 882679. GP acknowledges financial support from the European Research Council (ERC) under the European Union’s Horizon 2020 research and innovation program HotMilk (grant agreement No. 865637), support from Bando per il Finanziamento della Ricerca Fondamentale 2022 dell’Istituto Nazionale di Astrofisica (INAF): GO Large program and from the Framework per l’Attrazione e il Rafforzamento delle Eccellenze (FARE) per la ricerca in Italia (R20L5S39T9). LL acknowledges support from INAF grant 1.05.12.04.01. SE acknowledges financial contribution from the contracts Prin-MUR 2022 supported by Next Generation EU (M4.C2.1.1, n.20227RNLY3 {\it The concordance cosmological model: stress-tests with galaxy clusters}), and from the European Union’s Horizon 2020 Programme under the AHEAD2020 project (grant agreement n. 871158). {SVZ acknowledges support by the \emph{Deut\-sche For\-schungs\-ge\-mein\-schaft, DFG\/} project nr. 415510302.} BC acknowledges support provided by the Austrian Research Promotion Agency (FFG) and the Federal Ministry of the Republic of Austria for Climate Action, Environment, Energy, Mobility, Innovation and Technology (BMK) via the Austrian Space Applications Programme with grant numbers 899537 and 900565. EB acknowledges financial support from the European Research Council (ERC) Consolidator Grant under the European Union’s Horizon 2020 research and innovation program (grant agreement CoG DarkQuest No 101002585). {This research was supported by the Munich Institute for Astro-, Particle and BioPhysics (MIAPbP), which is funded by the Deutsche Forschungsgemeinschaft (DFG, German Research Foundation) under Germany´s Excellence Strategy – EXC-2094 – 390783311.} The calculations for the {\it {\sc Magneticum}} simulations were carried out at the Leibniz Supercomputer Center (LRZ) under project pr83li. This research has made using software provided by the Chandra X-ray Center (CXC) in the application packages CIAO and Sherpa. This work is based on data from eROSITA, the soft X-ray instrument aboard SRG, a joint Russian-German science mission supported by the Russian Space Agency (Roskosmos), in the interests of the Russian Academy of Sciences represented by its Space Research Institute (IKI), and the Deutsches Zentrum für Luft- und Raumfahrt (DLR). The SRG spacecraft was built by Lavochkin Association (NPOL) and its subcontractors, and is operated by NPOL with support from the Max Planck Institute for Extraterrestrial Physics (MPE). The development and construction of the eROSITA X-ray instrument was led by MPE, with contributions from the Dr. Karl Remeis Observatory Bamberg \& ECAP (FAU Erlangen-Nuernberg), the University of Hamburg Observatory, the Leibniz Institute for Astrophysics Potsdam (AIP), and the Institute for Astronomy and Astrophysics of the University of Tübingen, with the support of DLR and the Max Planck Society. The Argelander Institute for Astronomy of the University of Bonn and the Ludwig Maximilians Universität Munich also participated in the science preparation for eROSITA. The eROSITA data shown here were processed using the eSASS software system developed by the German eROSITA consortium.

\end{acknowledgements}

  \bibliographystyle{aa.bst} 
  \bibliography{aa54352-25.bib} 

\begin{appendix}

\section{Additional figures}

In this section, we present detailed results of the spectral analysis and the bootstrapping procedure applied to each mass bin. The halo mass and temperature values from different sources used in Figure \ref{fig:m_t} can be found in Table \ref{tab:values}. The stacked event lists, spectra for each individual mass bin, and histograms of the bootstrapping results for the observational sample can be found in Figure \ref{ap1}, and in Figure \ref{ap3} for the simulated sample.

\begin{figure}[h!]
   \centering
   \includegraphics[width=1\hsize]{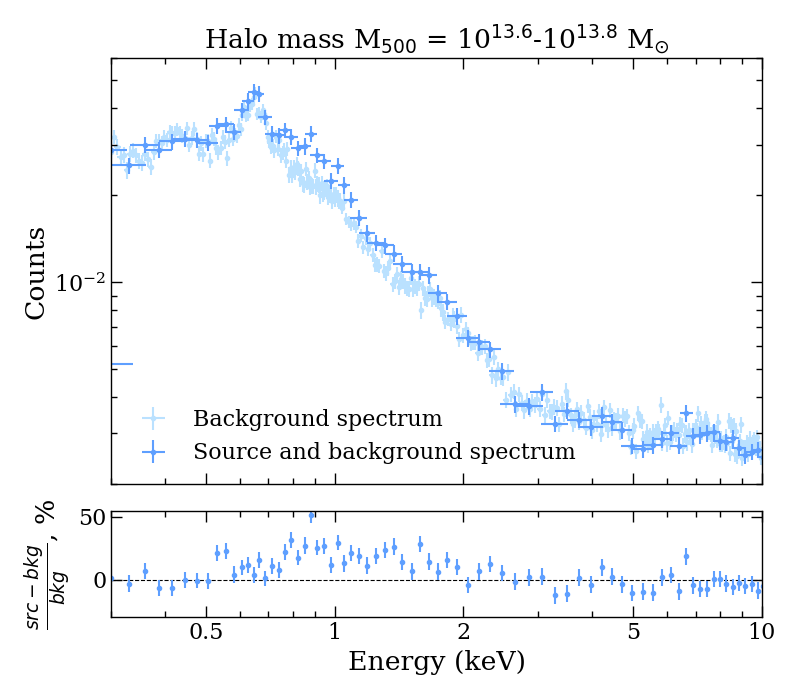}
      \caption{{Comparison of the source spectrum before background subtraction (dark blue) and the background spectrum (light blue) for one of the mass bins with masses ranging from $10^{13.6}$ to $10^{13.8} M_{\odot}$. The visible differences in the 0.7–2 keV energy range highlight the contribution of the source emission.} \label{fig:src_and_bkg}}
\end{figure}

{Finally, we confirm that the signal from the galaxy groups is clearly distinguishable from the background by comparing the background spectrum with the source spectrum prior to background subtraction. While the overall spectral shape is similar, Fig. \ref{fig:src_and_bkg} highlights a noticeable difference in the 0.7 -- 2 keV range. This excess emission is attributed to the stacked groups, confirming the detection of their contribution to the spectrum.}

\begin{table}[h!]
\caption{Spectral modeling results}   
\label{tab:values}
\centering
\renewcommand{\arraystretch}{1.5}
\begin{tabular}{|c|c|c|c|c|}
\hline
log$_{10}(M_{500})$ & kT (bkg subtraction) & kT (bkg modeling) \\
\small
$M_{500}$ in $M_{\odot}$ & keV & keV
\normalsize \\
\hline
\multicolumn{3}{|c|}{Observational data} \\
\hline
$13.09_{-0.09}^{+0.11}$ & $0.71_{-0.32}^{+0.16}$ & $0.71_{-0.41}^{+0.18}$ \\
$13.36_{-0.16}^{+0.24}$ & $0.89_{-0.08}^{+0.06}$ & $0.89_{-0.17}^{+0.23}$ \\
$13.76_{-0.16}^{+0.24}$ & $0.98_{-0.13}^{+0.08}$ & $0.99_{-0.12}^{+0.10}$ \\
$14.09_{-0.09}^{+0.11}$ & $2.47_{-0.81}^{+2.38}$ & $2.00_{-0.68}^{+1.58}$ \\
$14.31_{-0.11}^{+0.19}$ & $2.50_{-0.51}^{+0.96}$ & $2.78_{-0.89}^{+0.99}$ \\
$14.61_{-0.11}^{+0.19}$ & $5.00_{-0.86}^{+0.31}$ & $5.29_{-2.40}^{+1.51}$ \\
$14.85_{-0.05}^{+0.15}$ & $6.81_{-0.36}^{+2.62}$ & $7.41_{-2.88}^{+0.32}$ \\
\hline
\end{tabular}
\tablefoot{
Results of spectral stacking and modeling for groups in different mass bins presented in Figure \ref{fig:m_t}. The table lists the logarithmic mass, log$(M_{500}/M_{\odot})$, along with the corresponding temperatures, kT, derived from background subtraction and background modeling techniques. The values are presented with their 3$\sigma$ uncertainties.
}
\end{table}

\begin{figure*}[h!]
\centering
\includegraphics[width=1\hsize]{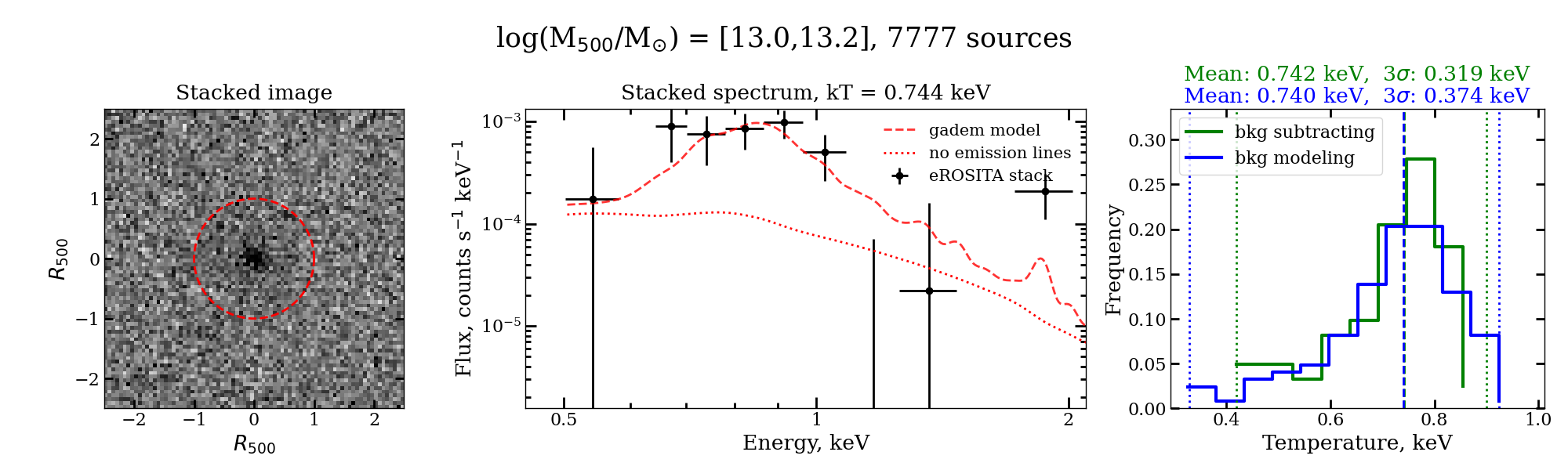}\\
\includegraphics[width=1\hsize]{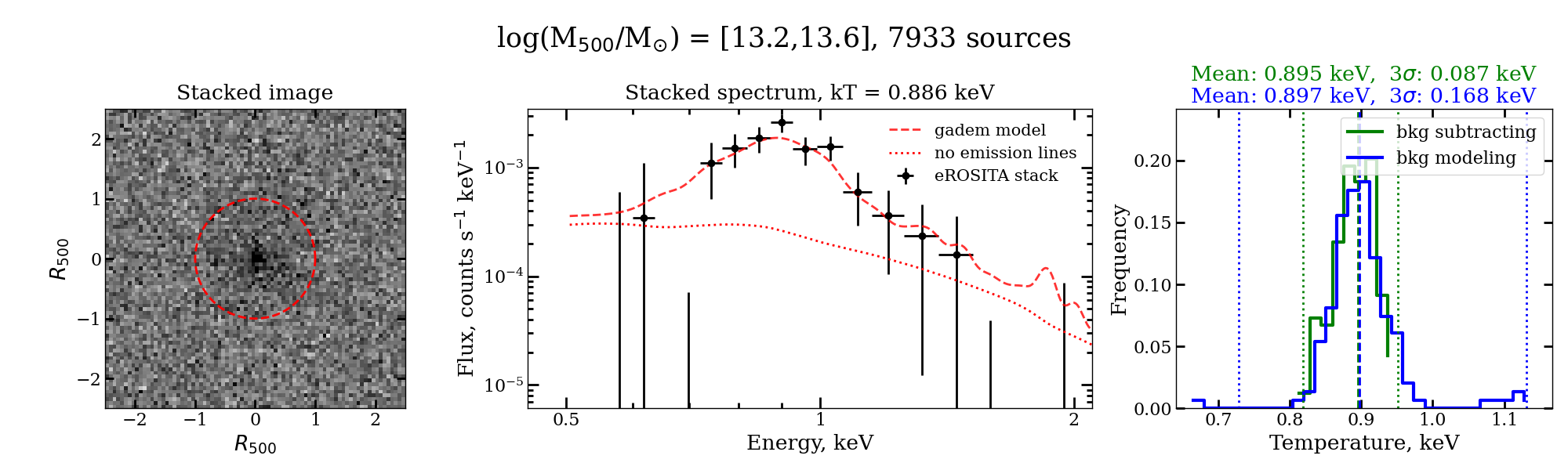}\\
\includegraphics[width=1\hsize]{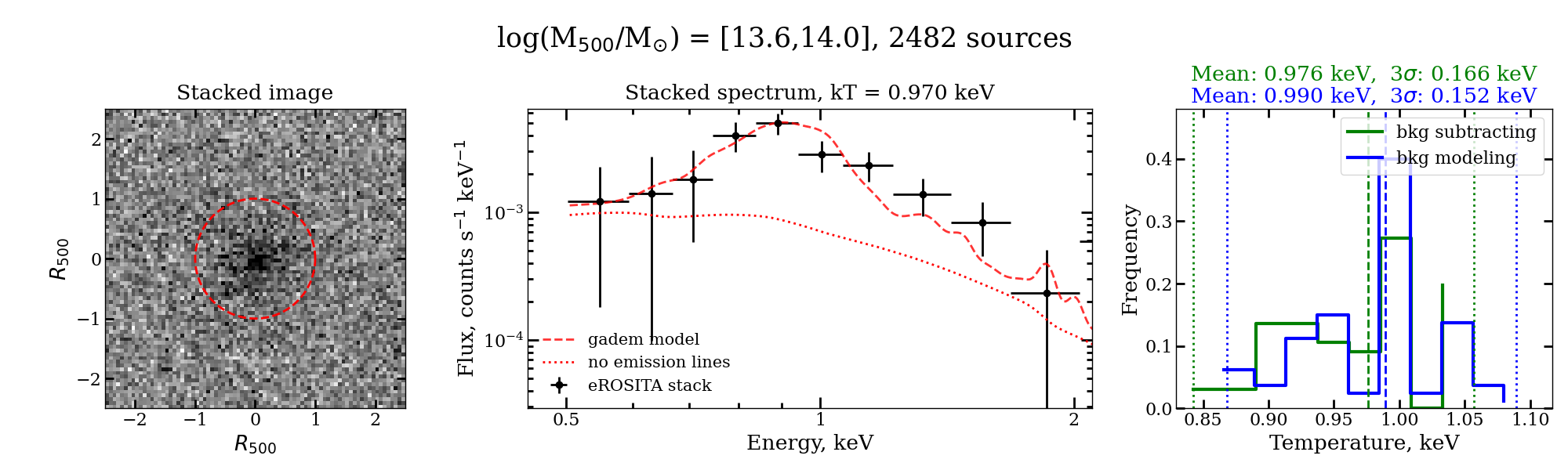}\\
\includegraphics[width=1\hsize]{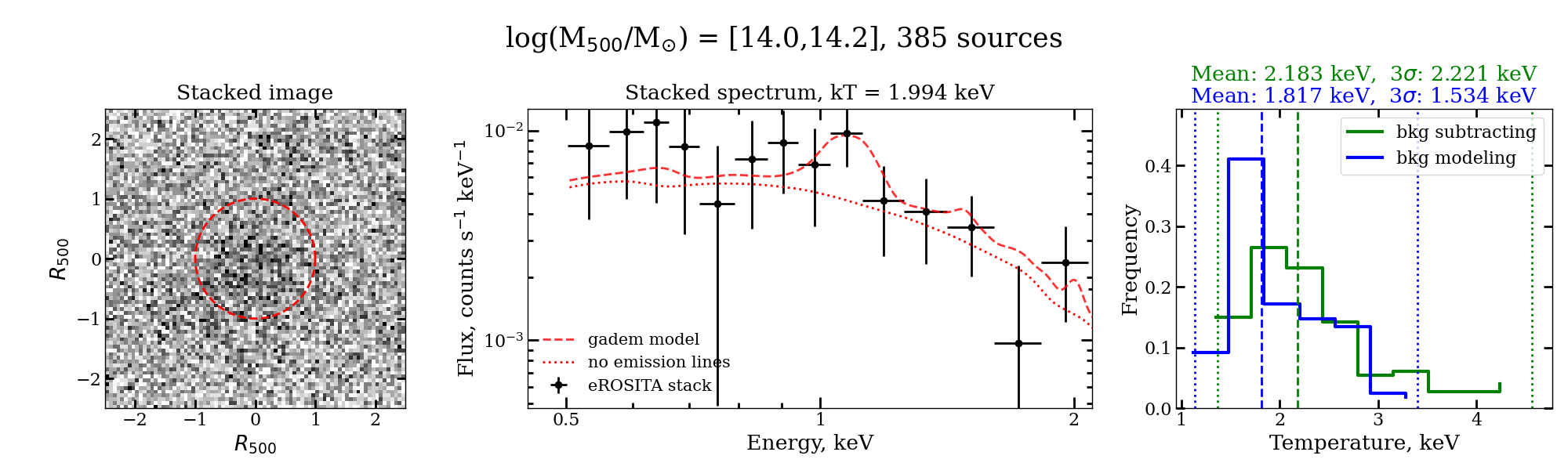}\\
\caption{\textit{Left panel}: Event lists scaled to $R_{500}$ and stacked on the coordinates of each source from the sample described in Section \ref{sec:description_obs}. The red circle indicates $R_{500}$. \textit{Central panel}: Stacked spectrum in the 0.5–2 keV energy range. The fit using {\sc gadem} with the mean temperature provided in the title is shown by the red dashed line. \textit{Right panel}: Histograms of temperatures from the bootstrapping analysis with background subtraction (green) and background modeling (blue). The title displays the mean temperature and the 3$\sigma$ deviation for the background modeling histogram.
\label{ap1}}
\end{figure*}

\begin{figure*}[h!]
\centering
\includegraphics[width=1\hsize]
{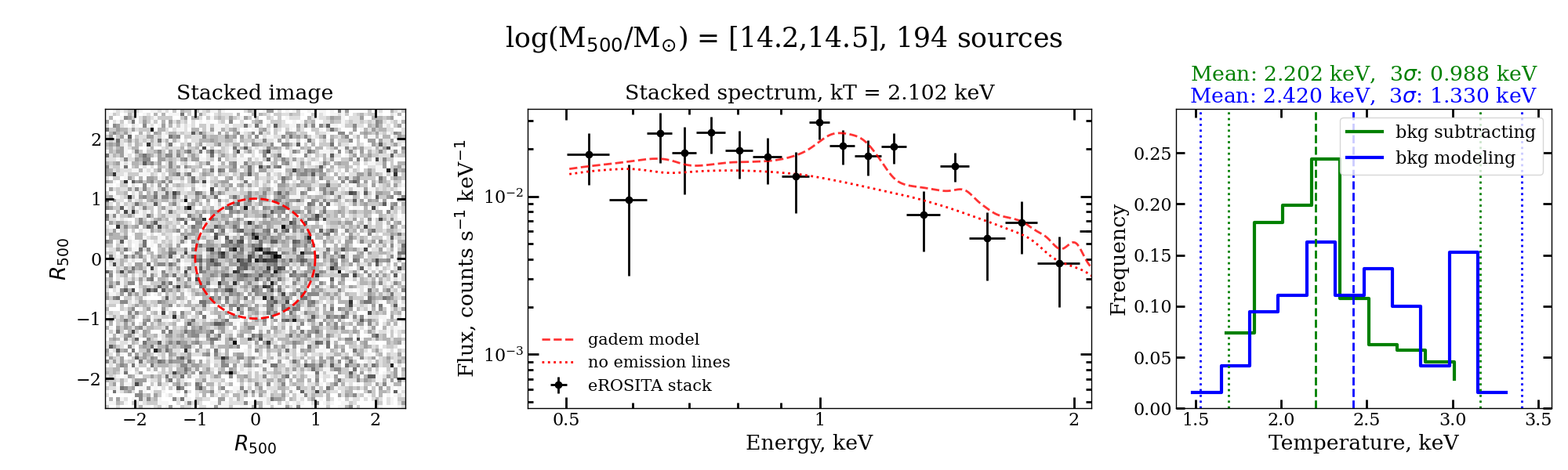}\\
\addtocounter{figure}{-1}
\caption{continued.}
\end{figure*}

\begin{figure*}
\centering
\includegraphics[width=1\hsize]{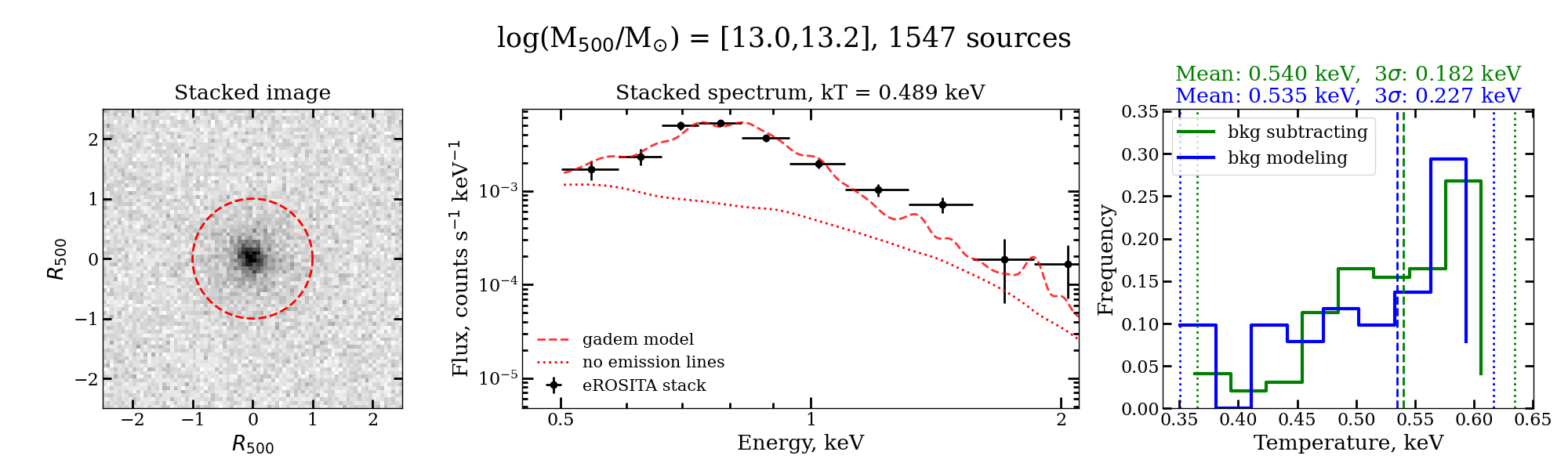}\\
\includegraphics[width=1\hsize]{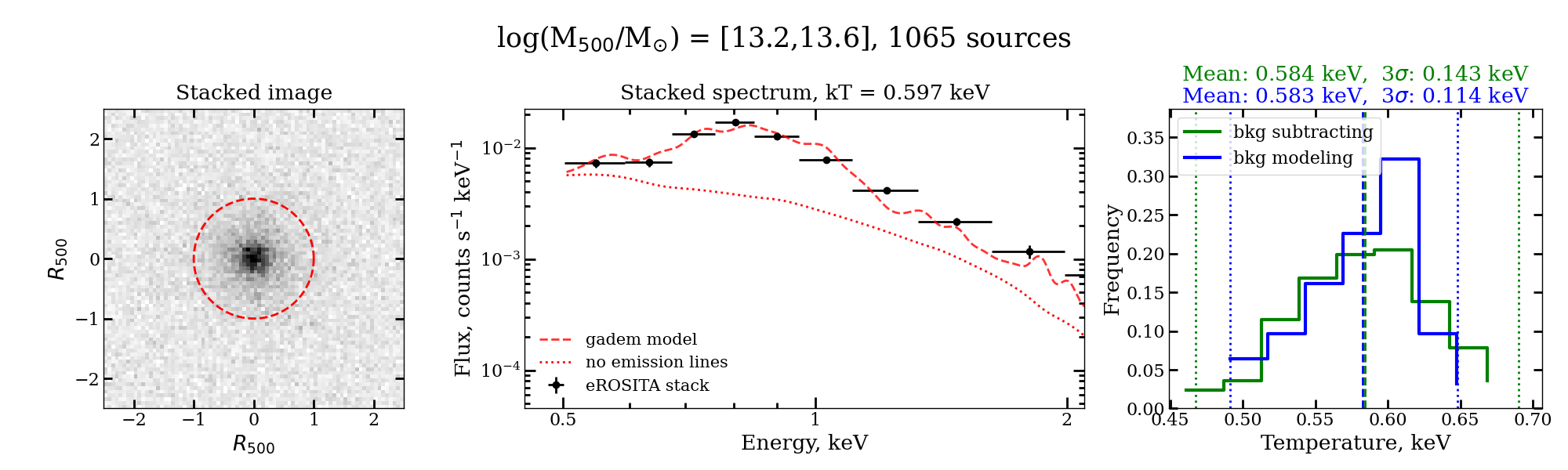}\\
\includegraphics[width=1\hsize]{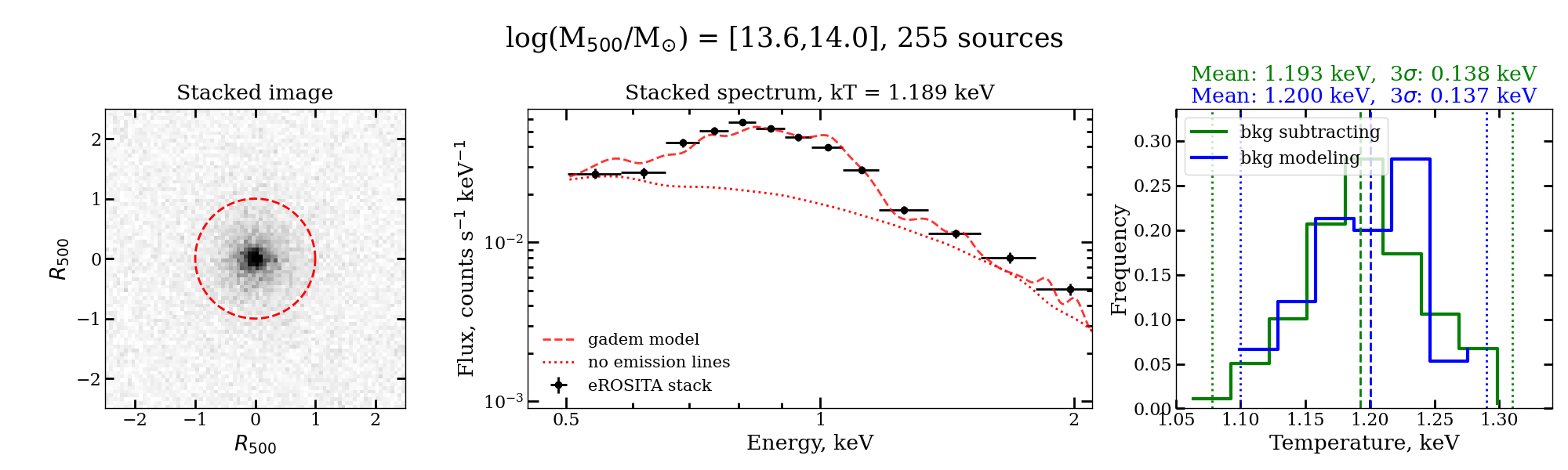}\\
\caption{Same as \ref{ap1} but for simulated sample described in \ref{sec:description_sims}.
\label{ap3}}
\end{figure*}

\twocolumn
\section{{Comparison of eRASS1 and eRASS4 in simulated data}}

{In this study, we used simulations based on eRASS4 depth, while the observational data correspond to eRASS1. The decision to use eRASS4 was motivated by two main reasons: (i) to obtain higher-quality spectra, which allow us to more precisely assess temperature reconstruction, and (ii) to explore how future eROSITA data releases might improve constraints on the $M-T$ relation.}

\begin{figure}[h!]
\centering
\includegraphics[width=0.49\hsize]{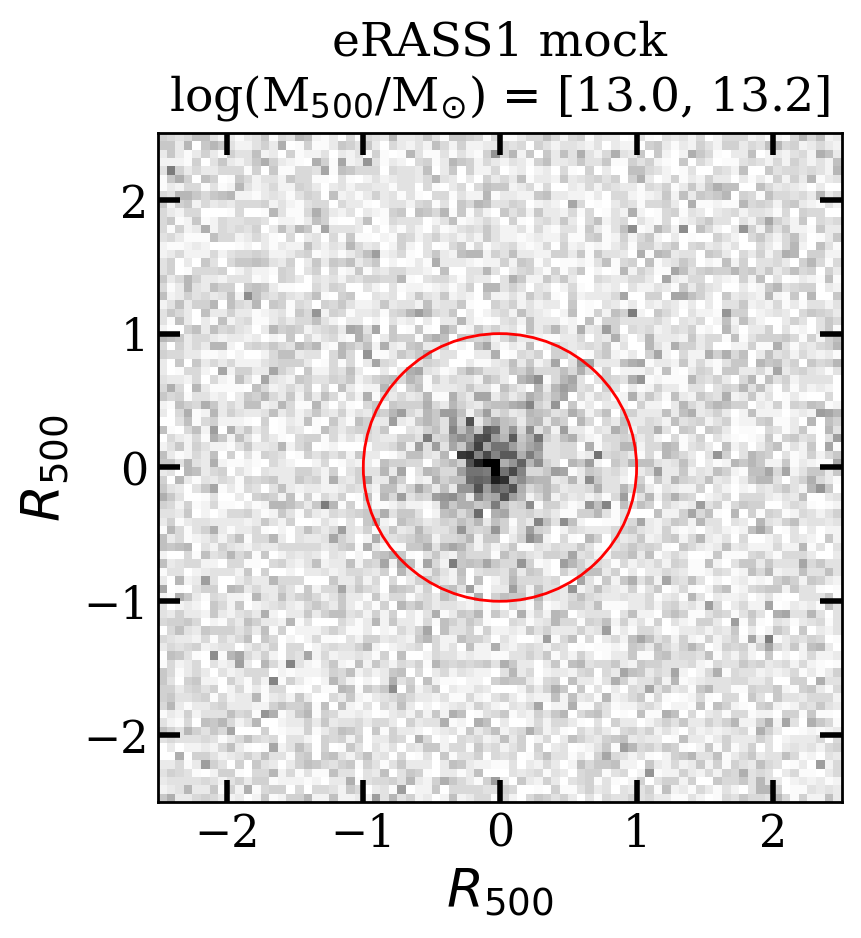}
\includegraphics[width=0.49\hsize]{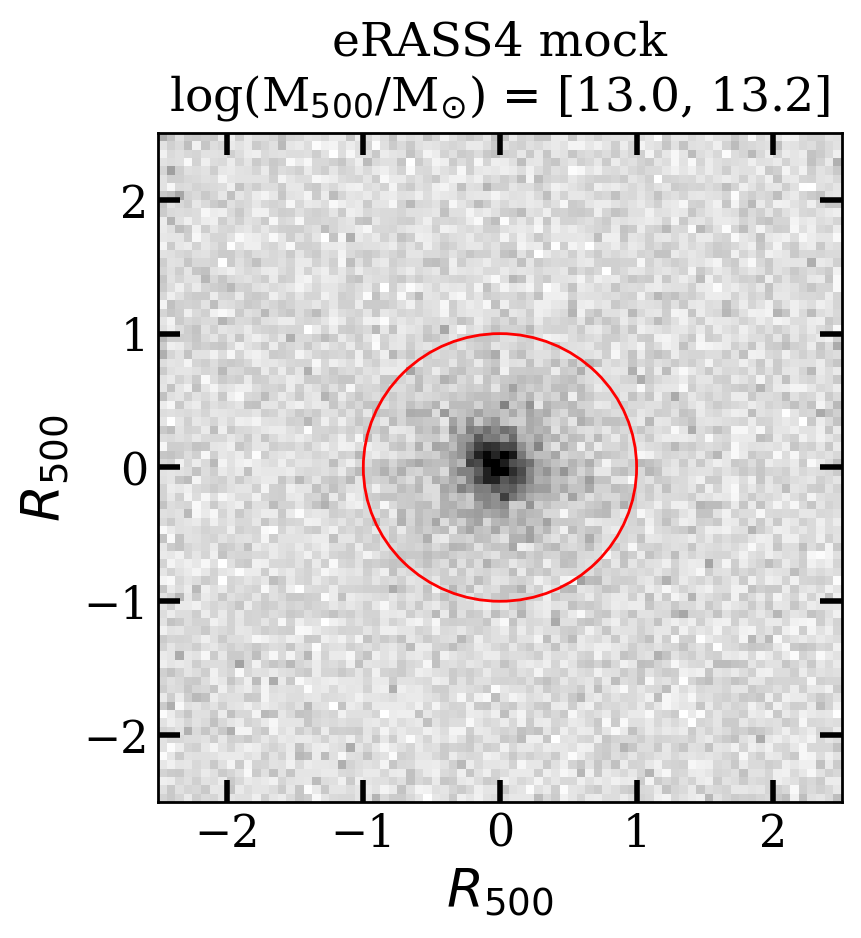}\\
\includegraphics[width=0.49\hsize]{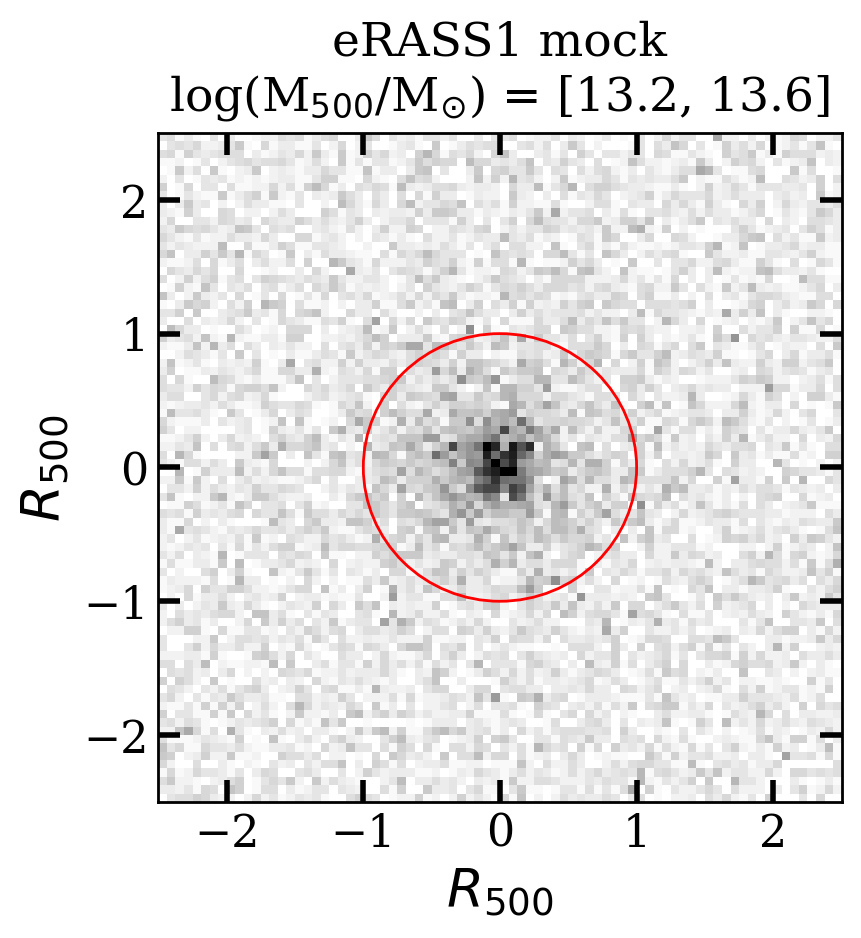}
\includegraphics[width=0.49\hsize]{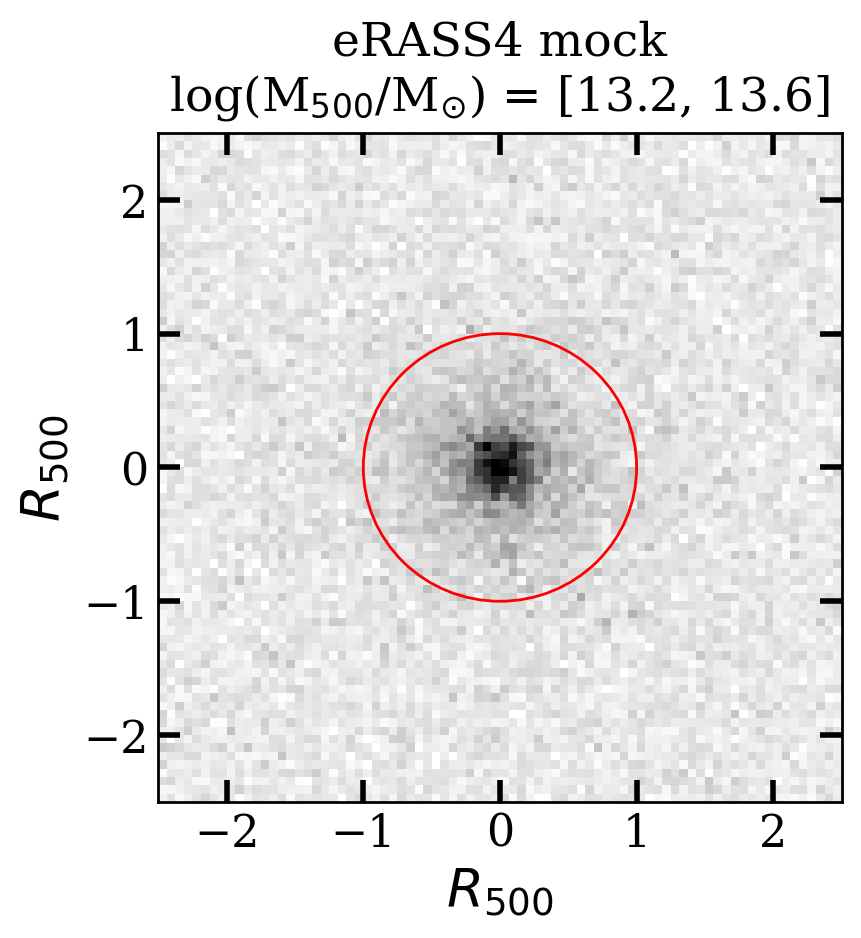}\\
\includegraphics[width=0.49\hsize]{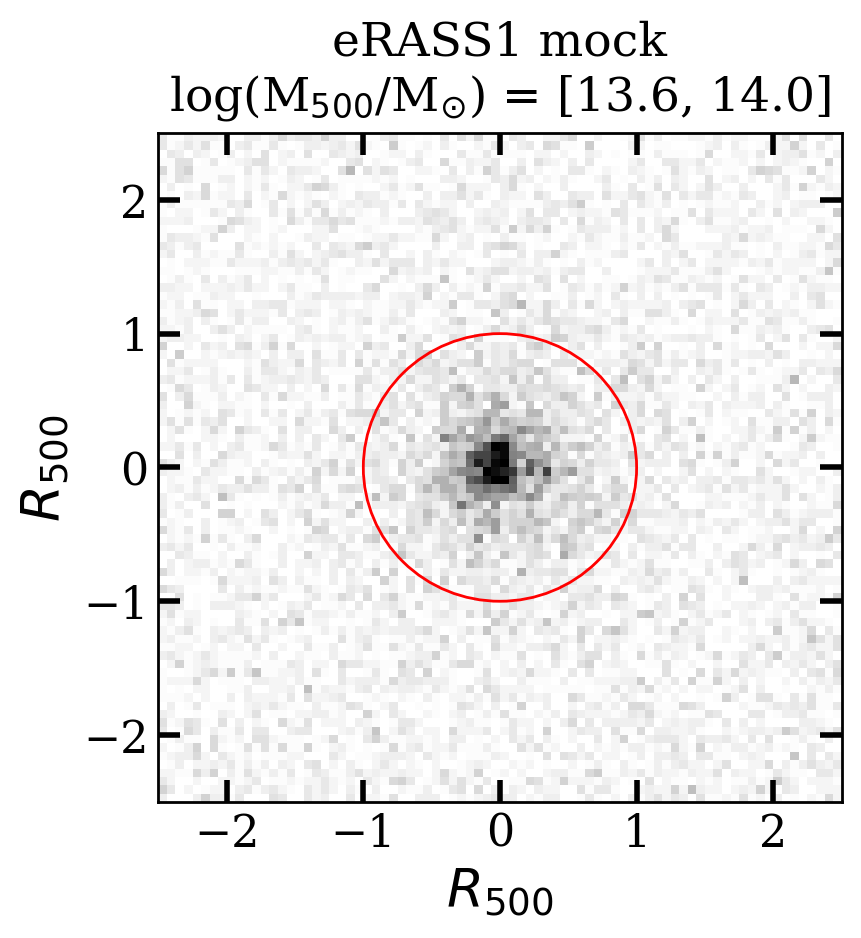}
\includegraphics[width=0.49\hsize]{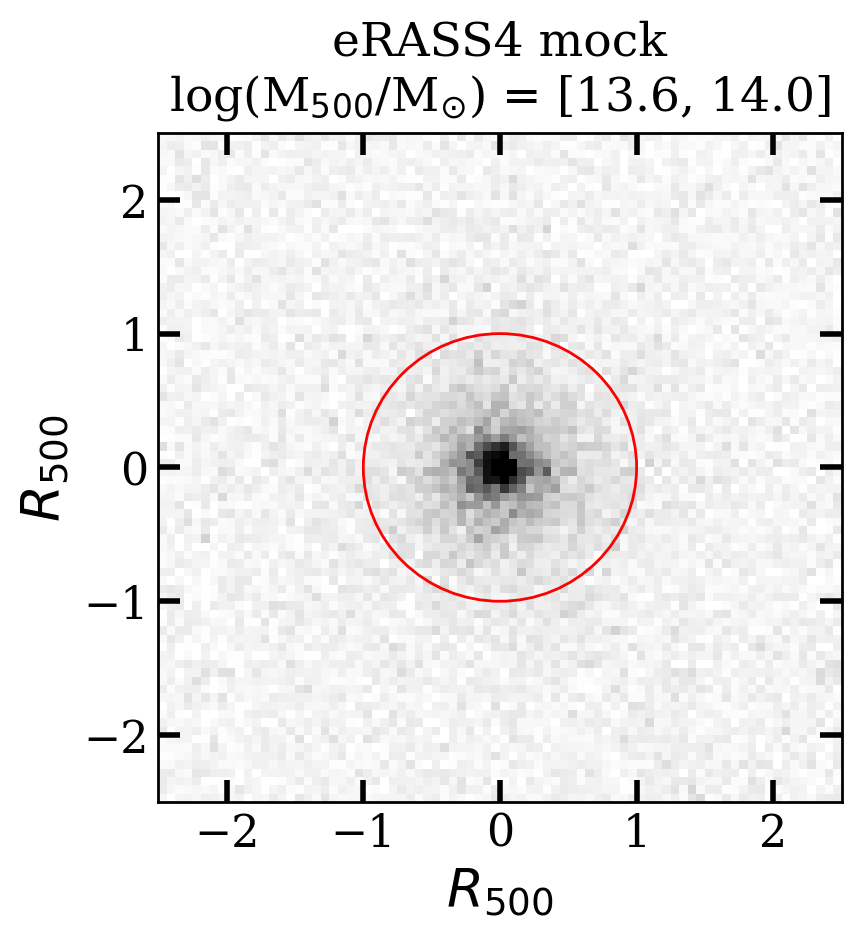}\\
\caption{{Event lists scaled to $R_{500}$ and stacked on the coordinates of each source from the sample described in Section \ref{sec:description_sims}. The red circle indicates $R_{500}$. The left and right columns show event lists corresponding to eRASS1 and eRASS4 depths, respectively.}
\label{apB}}
\end{figure}

{To ensure consistency and validate our approach, we also repeated the analysis using eRASS1 simulations. In this section, we provide both eRASS1 and eRASS4 simulated stacked images for selected mass bins for comparison. As shown in Figure~\ref{apB}, even the eRASS1 depth allows us to detect galaxy groups after stacking.}

\section{{Impact of halo center selection}}

{Another important assumption in our study is the choice of halo center used for stacking. We initially adopted the group centers from the \citet{yang_galaxy_2007} catalog -- geometrical, luminosity-weighted centers based on all group members identified by the Friends-of-Friends algorithm. However, these centers may not always coincide with the X-ray emission peak \citep{2016MNRAS.456.2566C}, especially if some group members were missed due to the limited depth of the SDSS survey.}

{To assess the effect of different centering choices, we compared the coordinates of the brightest cluster galaxy (BCG), also provided in the \citet{yang_galaxy_2007} catalog, with those of the group center. We found that in 74\% of the sample, the BCG is located within 0.2$R_{500}$ of the group center, and in 71\% of the sample, it is within 0.1$R_{500}$. However, larger offsets were observed in the remaining systems.}

{To investigate the impact of centering on the stacked images, we combined eROSITA event lists using the BCG coordinates instead of the group centers. The resulting images (Figure~\ref{apC}, left column) appear more centered and more closely resemble the simulated eRASS1 stacks (Figure~\ref{apB}, left column). This suggests that, for most systems in our sample, the X-ray emission peak is more closely aligned with the BCG than with the luminosity-weighted group center.}

{We also quantitatively evaluated the centering effect by comparing the number of ``source'' photons within $R_{500}$ for both centering choices. This was done by subtracting the photon counts in a background annulus of equal area from those within $R_{500}$. The difference was minimal, remaining within 0.54$\sigma$ across all mass bins. This indicates that the choice of halo center has a negligible impact on the total photon content within $R_{500}$ and, consequently, on our temperature estimates, although it may influence the spatial distribution of photons and the resulting light profiles.}

\begin{figure}[h!]
\centering
\includegraphics[width=0.49\hsize]{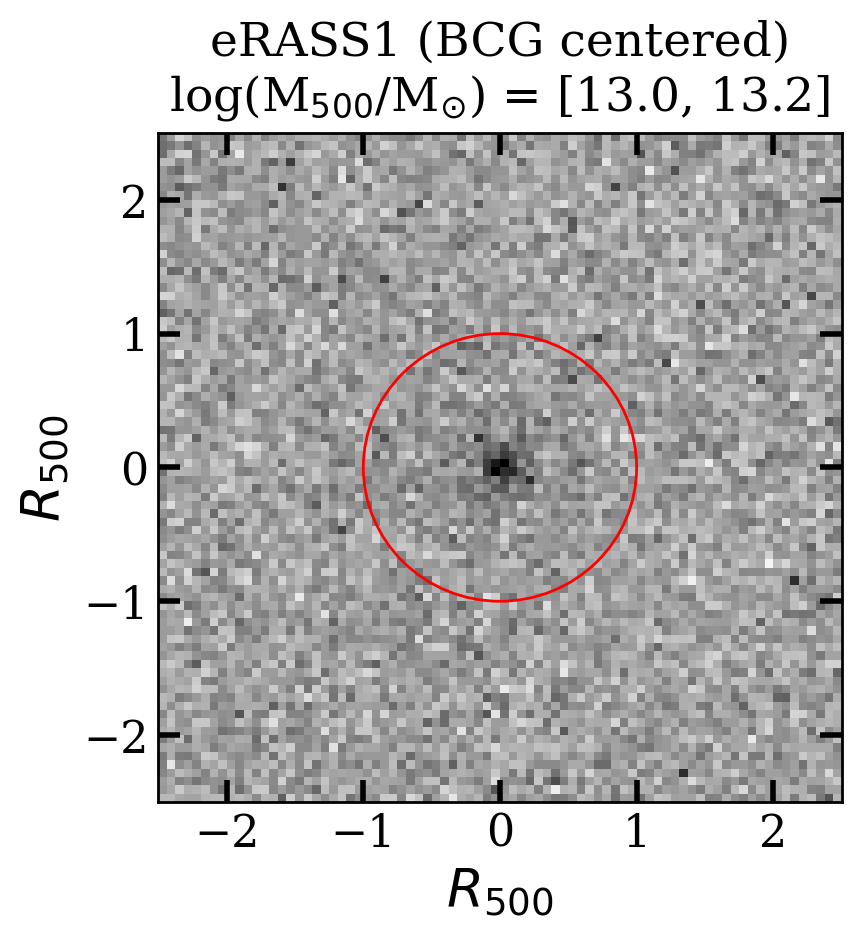}
\includegraphics[width=0.49\hsize]{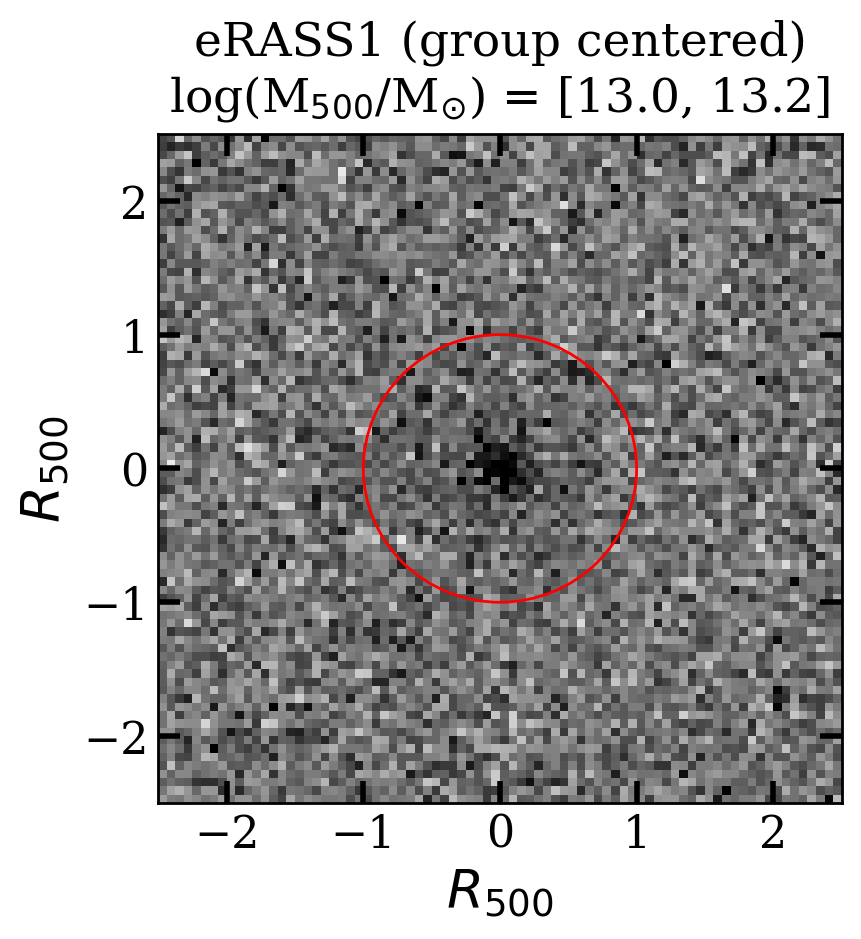}\\
\caption{{Event lists scaled to $R_{500}$ and stacked on the coordinates of BCG (left column) or light-weighted geometrical center (right column) of each source from the sample described in Section \ref{sec:description_obs}. The red circle indicates $R_{500}$.}
\label{apC}}
\end{figure}

\begin{figure*}
\centering
\includegraphics[width=0.24\hsize]{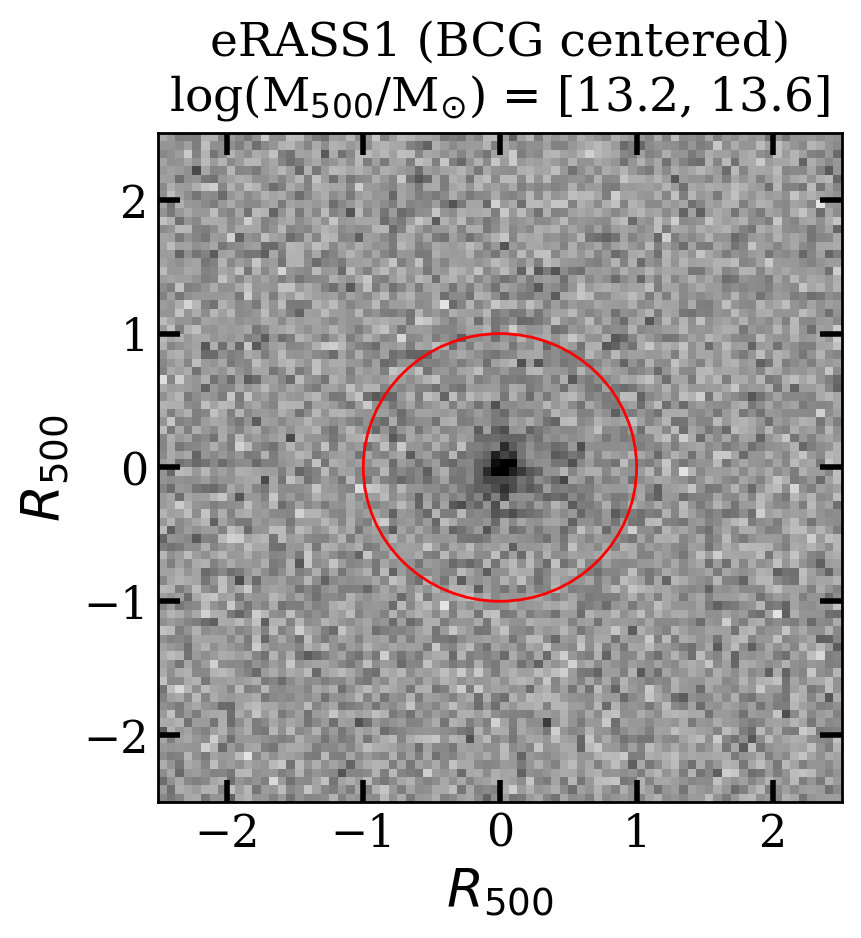}
\includegraphics[width=0.24\hsize]{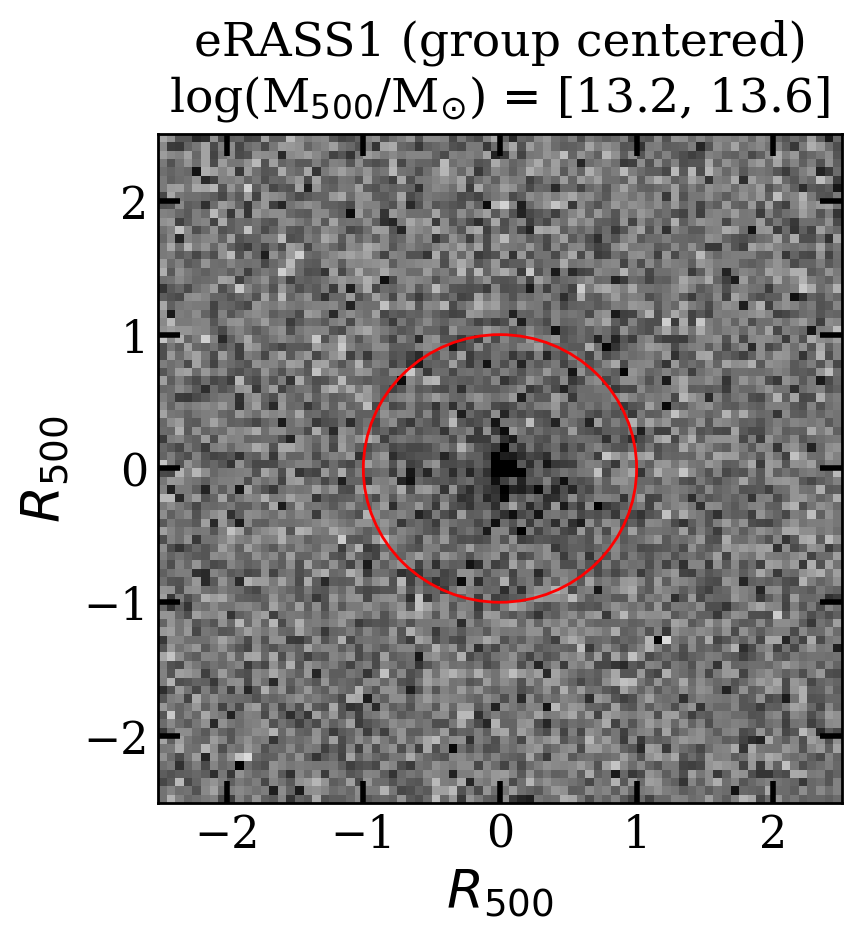}
\rule{0.4pt}{5cm}\hspace{0.01\hsize}
\includegraphics[width=0.24\hsize]{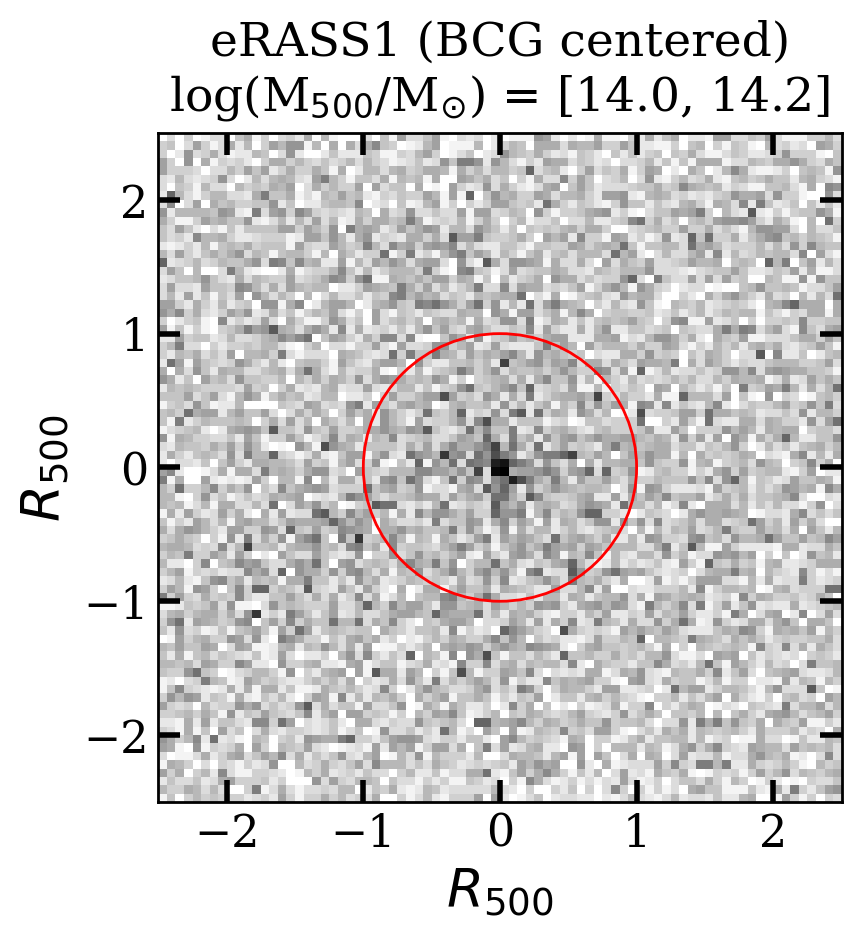}
\includegraphics[width=0.24\hsize]{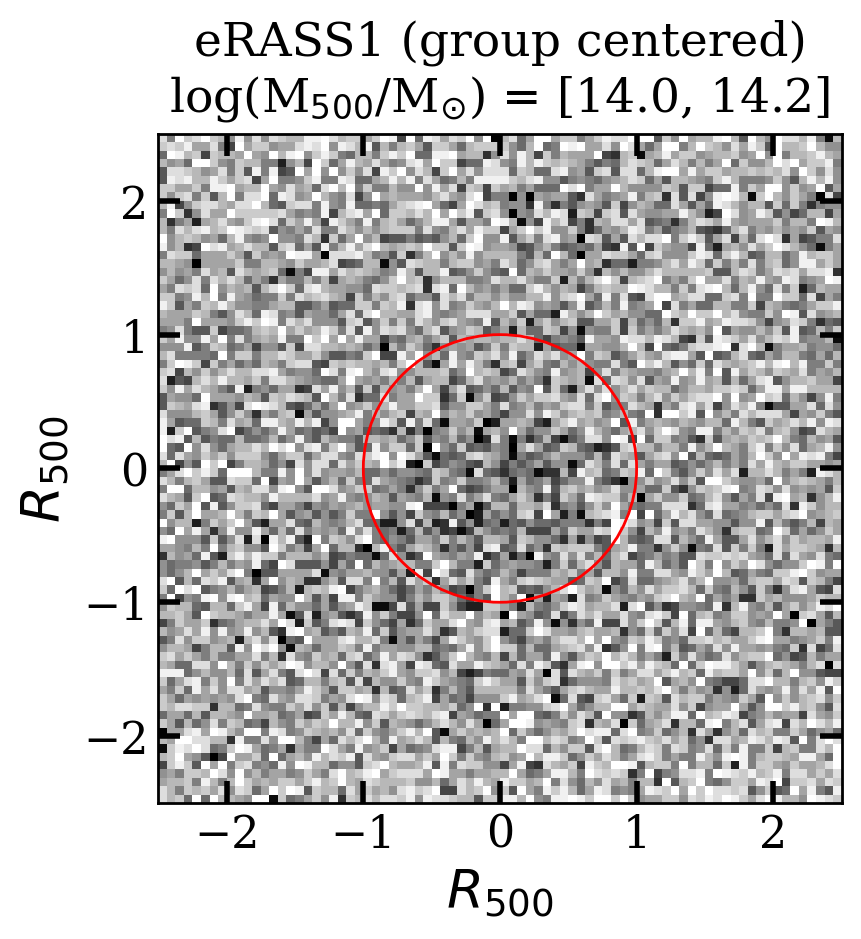}\\
\includegraphics[width=0.24\hsize]{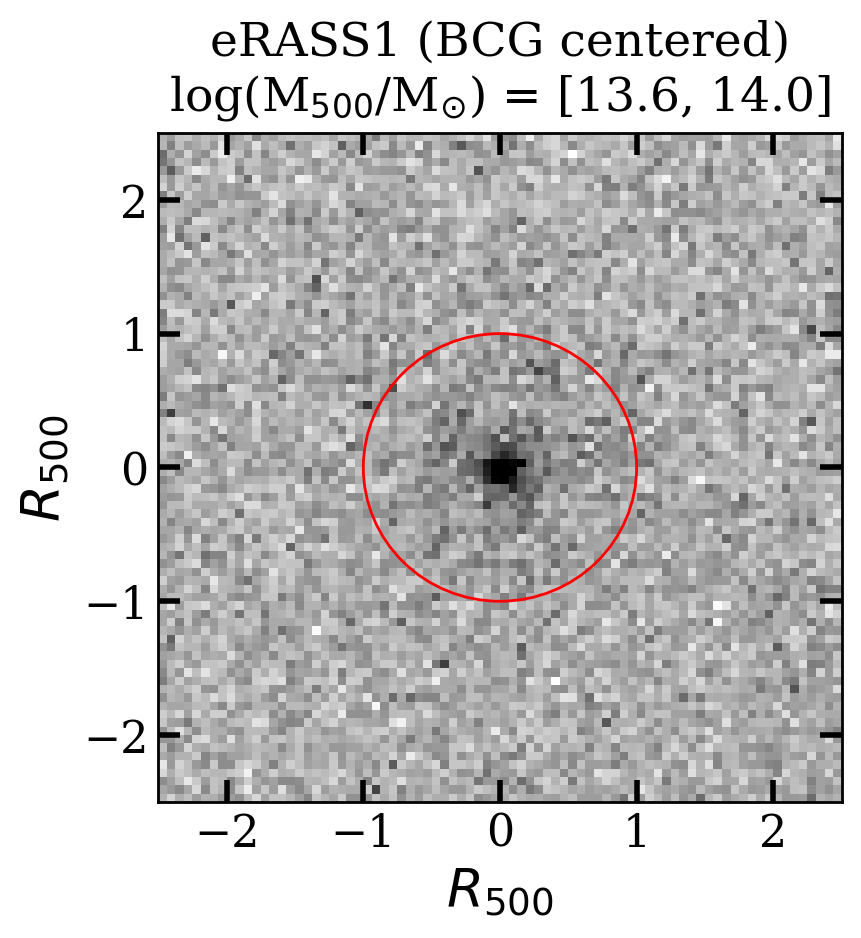}
\includegraphics[width=0.24\hsize]{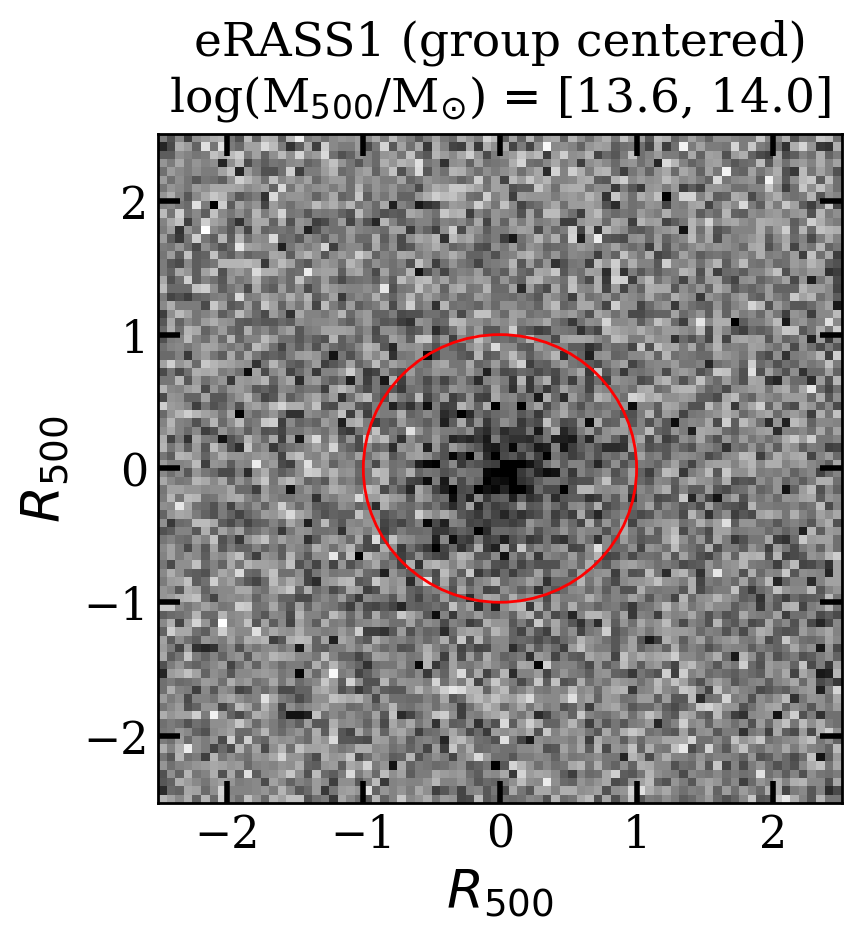}
\rule{0.4pt}{5cm}\hspace{0.01\hsize}
\includegraphics[width=0.24\hsize]{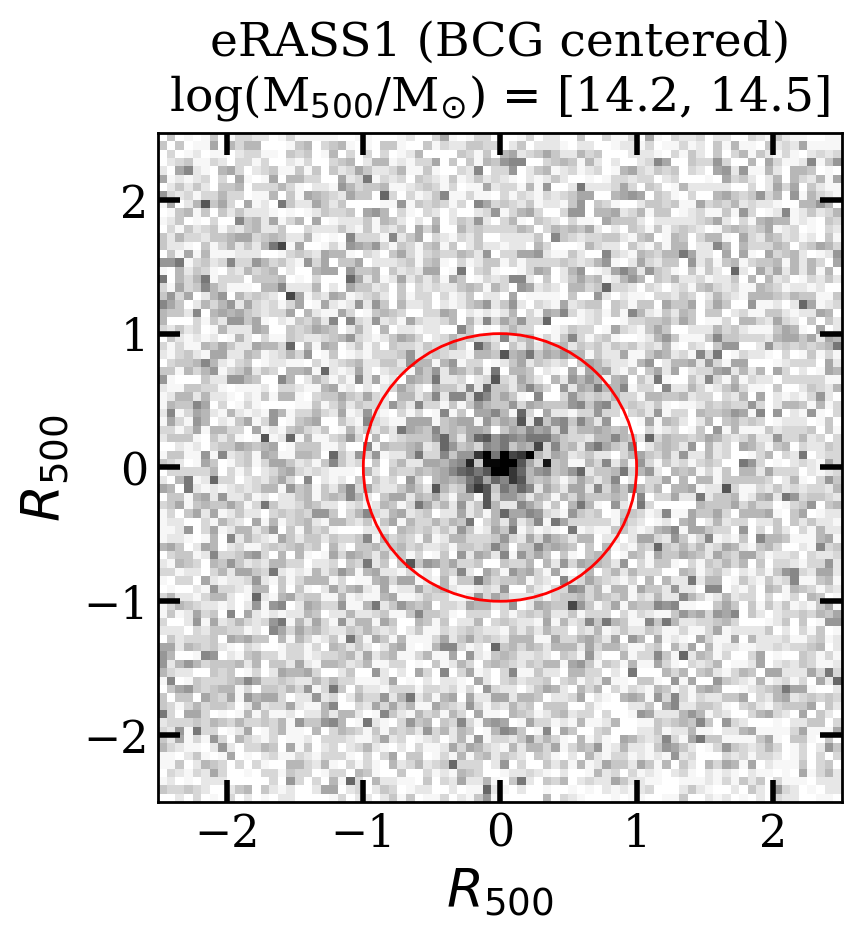}
\includegraphics[width=0.24\hsize]{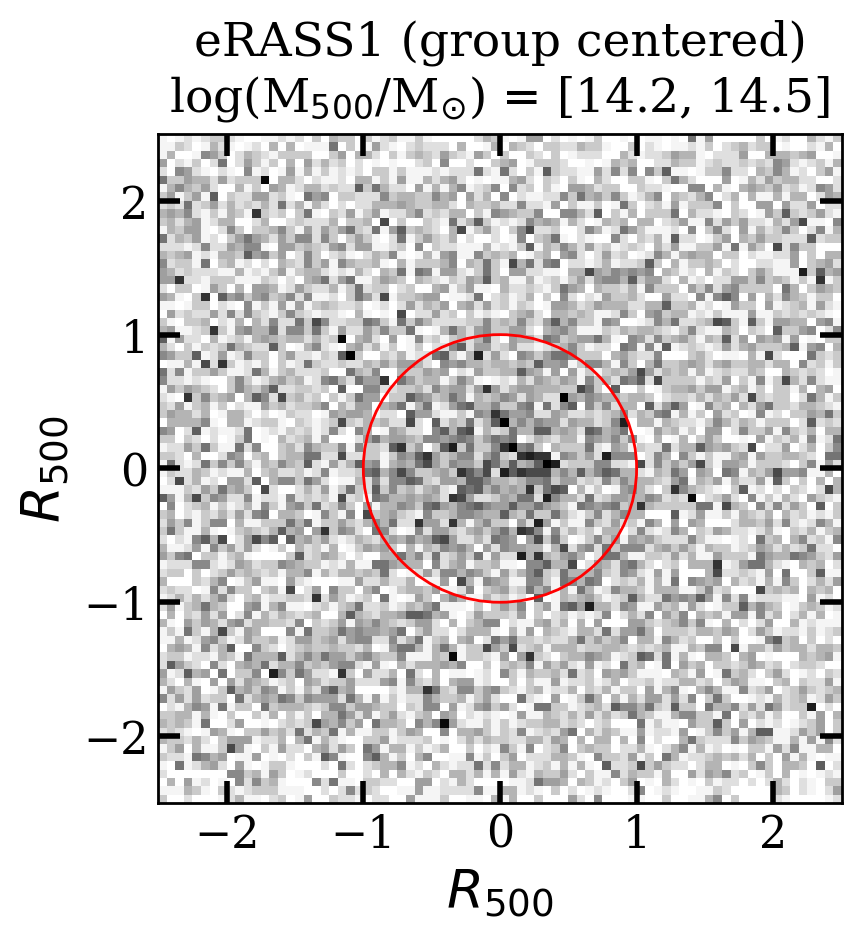}\\
\addtocounter{figure}{-1}
\caption{{continued.}}
\end{figure*}

\end{appendix}

\end{document}